\documentclass[structabstract]{aa}  
\usepackage{graphicx}
\usepackage[usenames,dvipsnames]{color}

\usepackage{txfonts}
\usepackage{longtable}
\usepackage{supertabular}
\usepackage{natbib}
\defcitealias{Fadda+08}{Paper 0}
\begin{document}

\newcommand{\ABi}[1]{\textcolor{magenta}{\bf Andrea says: #1}}

\newcommand{\mic}{$\mu$m$\,$}
\newcommand{\mica}{$\mu$m}
\newcommand{\lir}{$\rm{L}_{\rm{IR}} \,$}
\newcommand{\lira}{$\rm{L}_{\rm{IR}}$}
\newcommand{\tsfr}{$\Sigma$(SFR)$\,$}
\newcommand{\tsfra}{$\Sigma$(SFR)}
\newcommand{\tsfrm}{\Sigma$(SFR)$/M \,}
\newcommand{\tsfrma}{\Sigma$(SFR)$/M}
\newcommand{\sfrd}{{\rho}_{\rm{SFR}}}

\title{The role of massive halos in the Star Formation History of the Universe
\thanks{Herschel is an ESA space observatory with science instruments provided by 
European-led Principal Investigator consortia and with important 
participation from NASA.}}
\author{P. Popesso\inst{1,2}, A. Biviano\inst{3}, A. Finoguenov\inst{2}, D. Wilman\inst{2}, M. Salvato\inst{2}, B. Magnelli\inst{2}, C. Gruppioni\inst{4}, F. Pozzi\inst{5 }, G. Rodighiero\inst{6}, F. Ziparo\inst{2}, S. Berta\inst{2}, D. Elbaz\inst{7}, M.~Dickinson\inst{7},D. Lutz\inst{2}
\and B. Altieri\inst{9}
\and H. Aussel\inst{8}
\and A. Cimatti\inst{6}
\and D. Fadda\inst{10}
\and O. Ilbert\inst{11}
\and E. Le Floch\inst{8}
\and R. Nordon\inst{2}
\and A. Poglitsch\inst{2}
\and S. Genel\inst{12}
\and C.K. Xu\inst{13}
}
\offprints{Paola Popesso, popesso@mpe.mpg.de}

\institute{Excellence Cluster Universe, Boltzmannstr. 2, D-85748 Garching.
\and Max-Planck-Institut f\"{u}r Extraterrestrische Physik (MPE), Postfach 1312, 85741 Garching, Germany.
\and INAF/Osservatorio Astronomico di Trieste, via G.B. Tiepolo 11, I-34143 Trieste, Italy.
\and INAF-Osservatorio Astronomico di Bologna, via Ranzani 1, I-40127 Bologna, Italy.
\and Dipartimento di Astronomia, Universit{\`a} di Bologna, Via Ranzani 1, 40127 Bologna, Italy.
\and Dipartimento di Astronomia, Universit{\`a} di Padova, Vicolo dell'Osservatorio 3, 35122 Padova, Italy.
\and  National Optical Astronomy Observatory, 950 North Cherry Avenue, Tucson, AZ 85719, USA
\and Laboratoire AIM, CEA/DSM-CNRS-Universit{\'e} Paris Diderot, IRFU/Service
 d'Astrophysique,  B\^at.709, CEA-Saclay, 91191 Gif-sur-Yvette Cedex, France.
\and Herschel Science Centre, European Space Astronomy Centre, ESA, Villanueva de la Ca\~nada, 28691 Madrid, Spain
\and NASA Herschel Science Center, Caltech 100-22, Pasadena, CA 91125, USA
\and Institute for Astronomy 2680 Woodlawn Drive Honolulu, HI 96822-1897, USA 
\and Harvard-Smithsonian Center for Astrophysics, 60 Garden Street, MS-51, Cambridge, MA 02138 USA
\and IPAC, Caltech 100-22, Pasadena, CA 91125
}

\date{Received / Accepted}

\abstract{The most striking feature of the Cosmic Star Formation
  History (CSFH) of the Universe is a dramatic drop of the star
  formation (SF) activity, since $z \sim 1$.} {In this work we
  investigate if the very same process of assembly and growth of
  structures is one of the major drivers of the observed decline of
  the Universe SF activity.}{We study the contribution to the CSFH of
  galaxies in halos of different masses. This is done by studying the
  total SF rate-halo mass-redshift plane from redshift 0 to redshift
  $\sim$ 1.6 in a sample of 57 groups and clusters by using the
  deepest available mid- and far-infrared surveys conducted with
  Spitzer MIPS and Herschel PACS and SPIRE, on blank (ECDFS, CDFN and
  the COSMOS) and cluster fields.}{Our results show that low mass
  groups ($M_{halo} \sim 6{\times}10^{12}-6{\times}10^{13}$
  $M_{\odot}$) provide a 60-80\% contribution to the CSFH at
  z$\sim$1. Such contribution declines faster than the CSFH in the
  last 8 billion years to less than $10\%$ at $z<0.3$, where the
  overall SF activity is sustained by lower mass halos. More massive
  systems ($M_{halo} > 6{\times}10^{13}$ $M_{\odot}$) provide only a
  marginal contribution ($< 10\%$) at any epoch. A simplified abundance
  matching method shows that the large contribution of low mass groups
  at $z\sim 1$ is due to a large fraction ($>50\%$) of very massive,
  highly star forming Main Sequence galaxies. Below $z\sim 1$ a
  quenching process must take place in massive halos to cause the
  observed faster suppression of their SF activity. Such process must
  be a slow one though, as most of the models implementing a rapid
  quenching of the SF activity in accreting satellites significantly
  under-predicts the observed SF level in massive halos at any
  redshift. This would rule out short time-scale mechanisms such as
  ram pressure stripping. Instead, starvation or the satellite
  transition from cold to hot accretion would provide a quenching
  timescale of 1 to few Gyrs more consistent with the
  observations.}{Our results suggest a scenario in which, due to the
  structure formation process, more and more galaxies experience the
  group environment and the associated quenching process in the last 8
  billion years. This leads to the progressive suppression of their SF
  activity and, thus, it shapes the CSFH below $z\sim 1$.}


\titlerunning{The role of massive halos in the Star Formation History of the Universe}
\authorrunning{Popesso et al.}

\maketitle

\section{Introduction}
\label{s:intro}
Achieving an observational determination and a theoretical
understanding of the cosmic star formation history (CSFH) of the
Universe is still a big challenge in the study of galaxy formation. By
now, this history has been rather well established observationally up
to z$\sim$4
\citep{LeFloch+05,Perez-Gonzalez+2005,Caputi+07,reddy+08,magnelli+09,magnelli+11,magnelli+13,gruppioni+13}
and only sketched out to redshift z $\sim$ 6-7 with larger
uncertainties. The most striking feature of the CSFH, suggested by
essentially all star-formation (SF) activity indicators, is that the
star-formation rate (SFR) per unit volume in the Universe was an order
of magnitude greater at z$\sim$1 than in the present day
\citep{Lilly+96,Madau+98,LeFloch+05,magnelli+09} and that the SF
density stays at comparable or even higher levels out to at least
redshift z $\sim$ 2-3 \citep{reddy+08,soifer+08,Hopkins_Beacom06}. The
analysis of the contribution of different classes of galaxies to the
CSFH revealed another interesting aspect. The contribution of highly
star forming galaxies (Luminous Infrared Galaxies, LIRGs, $L_{IR} >
10^{11} L_{\odot}$), although negligible in the local Universe,
becomes comparable to that of normal star forming galaxies around
z$\sim$1, and they dominate during the whole active phase at
z$\sim$1-3 \citep{LeFloch+05,magnelli+09}. The most powerful starburst (SB)
galaxies (Ultra-luminous Infrared Galaxies, ULIRGs, $L_{IR} > 10^{12}
L_{\odot}$) undergo the fastest evolution dominating the CSFH only at
z$\sim$2 and 3 and disappearing, then, by redshift $\sim$0
\citep{cowie+04}. Given the existence of the so called "Main Sequence"
(MS) of star forming galaxies, the galaxy SF activity is tightly
linked to the galaxy stellar mass. This relation holds from redshift
$\sim$0 up to redshift $\sim$2 with a rather small dispersion (0.2-0.3
dex) and with normalization monotonically increasing with redshift
\citep{Noeske+07,Elbaz+07,Daddi+07,Peng+10}.  This relation supports
the so called "galaxy downsizing" scenario. Namely most massive
galaxies seem to have formed their stars early in cosmic history and
their contribution to the CSFH was significantly larger at higher
redshifts through a very powerful phase of star formation activity
(LIRGs, ULIRGs and sub-mm galaxies). Low mass galaxies seem to have
formed much later and they dominate the present epoch through a mild
and steady SF activity (fainter infrared galaxies).

The most obvious reason for a galaxy to stop forming stars is the lack
of gas supply. Indeed, high-z galaxies show a larger gas content with
respect to the present star forming systems \citep{tacconi+10}. The
most accredited models of galaxy formation advocate Active Galactic
Nuclei (AGN) feedback as the main mechanism to drive the gas away and
stop the growth of the galaxy and its central black hole (BH). These
models are able to explain at the same time the observed drop in the
CSFH and the correlation of BH and host galaxy masses
\citep{magorrian+98}. However, observations have difficulty in finding
evidence for the existence of such a feedback for normal galaxies
\citep{rovilos+12,mullaney+12,bongiorno+12,rosario+12,harrison+12}.
      Observations of the local Universe show that the BH growth
          is switched on with a delay with respect to the SB
          phase and that it is fueled by recycled gas from inner bulge
          stars \citep{schawinski+09,wild+10,yesuf+14}. These results
          are supported by models showing that the feedback from SF
          itself is at least as strong as that from an AGN; thus, if
          SF is in need of being quenched, AGN feedback generally does
          not play the primary role \citep{cen+12}.  Thus, a
      different quenching process, or a combination of many of them,
      perhaps including AGN feedback, is required to explain
      observations \citep{Peng+10}. Alternative candidates for
      quenching are those processes related to the environment, like,
      e.g. ram pressure stripping and gas starvation. These processes
      are often invoked to explain why galaxies in nearby groups and
      clusters are different than those in the field, in terms of morphology
      \citep[the morphology-density relation][]{dressler+80}), gas
      content \citep[the HI gas deficiency][]{Gavazzi+06,verdes+01},
      and SF activity, \citep[the SFR-density
        relation,][]{Gomez+03,popesso+12}.

Since the number density of groups and clusters with
$M>10^{12.5-13}M_{\odot}$ was a factor 10 lower at z$\sim$1 than now
\citep{williams+12}, an increasingly larger fraction of galaxies
\citep[60-70\% at z$\sim$0,][]{eke+05} has experienced the group
environment with cosmic time. The late-time growth of group-sized
halos occurs in parallel to the progressive decline of the SF activity
of the Universe since z$\sim$1.  Thus, if the group environment is the
site of physical processes that quench the star formation activity,
the very same process of assembly and growth of structures may be at the
origin of the strong decline in the CSFH.

The most straightforward way to explore this possibility is to follow
the approach of investigating the contribution of different classes of
galaxies to the CSFH, focusing the analysis not on individual
galaxies (as generally done) but on their parent halos. Indeed, if it
is the environment that drives the evolution of the star formation
activity, then we should really be classifying galaxies based on the
parent halo mass. Thus, in this paper we provide the first attempt of
measuring the differential contributions to the CSFH of galaxies
within Dark Matter (DM) halos of different masses. The two main
ingredients required to perform such an investigation are the
knowledge of the evolution of the SFR distribution of galaxies in DM
halos of different masses, and a way to classify galaxies according to
their parent halo mass.

The first ingredient is provided by the analysis of the evolution of
the Infrared (IR) Luminosity function (LF) of group and cluster
galaxies with respect to more isolated field galaxies. This aspect is
investigated in detail in a companion paper \citep{popesso+14}.

The second ingredient, the parent halo mass, is not an observable and,
therefore, the detection and selection methods of ``halos'' have to be
based on other group/cluster observable properties. Galaxy clusters
and groups are permeated by a thin hot intracluster medium, compressed
and shock heated during the halo collapse to temperatures $\sim 10^7$
keV and radiating optically thin thermal bremsstrahlung radiation in
the X-ray band. The X-ray selection is thus the best mean to select
galaxy groups and clusters and to avoid wrong galaxy groups
identifications due to projection effects typical of the optical and
lensing selection techniques.  Under the condition of hydrostatic
equilibrium, the gas temperature and density are directly related to
halo mass. A tight relation (rms $\sim$0.15 dex) exists also between
the cluster dynamical mass and the X-ray luminosity
\citep[$L_X$,][]{pratt+07,rykoff+08}. Even though this relation shows
a larger scatter for groups \citep[rms $\sim$0.3
  dex,][]{sun12,Leauthaud+10}, it is sufficiently tight to allow classifying
galaxies in parent-halo mass bins of $\sim$0.5 dex.

In this paper we use the analysis done in Popesso et al. (2014) about
the evolution of the IR group and cluster LF from redshift 0 to
redshift $z \sim 1.6$ to investigate the evolution of the relation
between the DM halo total SFR and the host halo mass in a rather large
sample of X-ray selected groups and clusters. We use the newest and
deepest available mid- and far-infrared surveys conducted with
{\it{Spitzer}} MIPS and with the most recent Photodetector Array
Camera and Spectrometer (PACS) on board of the {\it{Herschel}}
satellite, on the major blank fields such as the Extended Chandra Deep
Fields South (ECDFS), the Chandra Deep Field North (CDFN) and the
COSMOS field. Indeed, all these fields are part of the largest GT and
KT Herschel Programmes conducted with PACS: the PACS Evolutionary
Probe \citep{Lutz+11} and the GOODS-Herschel Program
\citep{Elbaz+11}. In addition, the blank fields considered in this
work are observed extensively in the X-ray with $Chandra$ and
$XMM-Newton$. The ECDFS, CDFN and COSMOS fields are also the site of
extensive spectroscopic campaigns that led to a superb spectroscopic
coverage. This is essential for the identification of group members
using the galaxy redshifts (and positions). The evolution of the IR
group LF is used to study the SFR distribution of group galaxies and
to measure their contribution to the CSFH.

The paper is structured as follows. In Sect. \ref{s:data} we describe
our data-set. In Sect. \ref{estimate} we describe the method used to
estimate the total SFR and the total SFR per unit of halo mass
group. In Sect.  \ref{sfr} we analyze how the relation between total
DM halo SF activity and host halo mass evolves with redshift. In
Sect. \ref{s:sfr_density} we use this analysis and the predicted
evolution of the DM halos comoving number density to reconstruct the
contribution to the CSFH of galaxy populations inhabiting halos of
different masses. In Sect. \ref{comparison} we compare our results
with the predictions of different types of theoretical models. In
Sect.  \ref{conclusion} we draw our conclusions.  We adopt H$_0=70$
km~s$^{-1}$~Mpc$^{-1}$, $\Omega_m=0.3$, $\Omega_{\Lambda}=0.7$
throughout this paper.

\section{The data-set}
\label{s:data}
The baseline for our analysis is provided by the galaxy group sample
described in Popesso et al. (2014). In the following section we
briefly describe this data-set and how we complement this group sample
with additional lower redshift groups and with galaxy clusters to
fully cover the redshift range from 0 to $\sim$1.6 and the full
dynamical range of massive halos with $M_{halo} > 10^{12.5-13}
M_{\odot}$.

The galaxy group sample of Popesso et al. (2014) comprises the X-ray
selected group sample of \cite{popesso+12} drawn from the X-ray galaxy
group catalog of COSMOS and CDFN, and the X-ray selected group sample
of \cite{ziparo+13} drawn from the X-ray group catalog of CDFS. All
catalogs are derived either from $Chandra$ or XMM-$Newton$
observations of such fields following the data reduction of Finoguenov
et al. (2014, in prep.).

All considered fields are covered by deep observations with
{\it{Spitzer}} MIPS at 24 $\mu$m and {\it{Herschel}} PACS at 100 and
160 $\mu$m. For COSMOS the source catalogs are taken from the public
data releases of Spitzer 24 \mic \citep{LeFloch+09,Sanders+07} and PEP
PACS 100 and 160 \mic \citep{Lutz+11,magnelli+13}.  For CDFN and CDFS,
and the inner GOODS regions, the source catalogs are taken from the
{\it{Spitzer}} MIPS 24 $\mu$m Fidel Program \citep{magnelli+09} and
from the combination of the PACS PEP \citep{Lutz+11} and
GOODS-Herschel \citep{Elbaz+11} surveys at 70, 100 and 160 $\mu$m
\citep{magnelli+13}. The reader is referred to Popesso et al. (2014)
for the details about the flux limits of each survey.

The association between 24 \mic and PACS sources with their optical
counterparts (the optical catalog of \citealt{Capak+07},
\citealt{cardamone+10} and \citealt{Berta+10} for COSMOS, CDFS and
CDFN, respectively) is done via a maximum likelihood method
\citep[see][for details]{Lutz+11}. The photometric sources were
cross-matched in coordinates with the available catalogs of
spectroscopic redshifts. For COSMOS this redshift catalog comes from
either SDSS or the public zCOSMOS-bright data acquired using VLT/VIMOS
\citep{Lilly+07,Lilly+09}, and is complemented with Keck/DEIMOS (PIs:
Scoville, Capak, Salvato, Sanders, Kartaltepe), Magellan/IMACS
\citep{Trump+07}, and MMT \citep{Prescott+06} spectroscopic
redshifts. For CDFS the redshift compilation includes the redshift
catalogs of \citet{cardamone+10} and \cite{silverman+10}, as well as
the redshift catalogs of the Arizona CDFS Environment Survey
\citep[ACES,][]{cooper+12}, and the GMASS survey
\citep{Cimatti+08}. The reader is referred to Popesso et al. (2014)
for a detailed discussion about the spectroscopic completeness as a
function of the {\it{Spitzer}} MIPS 24 $\mu$m flux. For CDFN we use the redshift compilation of \cite{barger+08}.

We use the spectroscopic information to define the group membership of
each system through the use of the {\it Clean} algorithm of
\citet{Mamon+13}, which is based on the modeling of the mass and
anisotropy profiles of cluster-sized halos extracted from a
cosmological numerical simulation. The procedure is iterated until it
finds a stable solution for the group velocity dispersion and, thus,
the group membership.

As explained in Popesso et al. (2014), from the initial COSMOS, CDFN
and CDFS X-ray group catalogs, only the groups with more than 10
members and located in a region of high spectroscopic coverage ($>
60\%$ at fluxes higher than 60 $\mu$Jy in the {\it{Spitzer}} 24 $\mu$m
band) are retained in the final sample. The final group sample
comprises 39 groups. We stress that a minimum of 10
spectroscopic members are required for a secure velocity dispersion
measurement and, thus, a secure membership definition. This selection
does not lead to a bias towards rich systems in our case. Indeed,
there is no magnitude or stellar mass limit imposed to the required 10
members. Thus, the very high spectroscopic completeness, in
particular of CDFN and CDFS \citep[see][for
  details]{Popesso+09,cooper+12}, leads to the selection of faint and
very low mass galaxy groups. Thus, if the group richness is defined as
the number of galaxies brighter than a fixed absolute magnitude limit
or more massive than a stellar mass limit, our sample is covering a
very broad range of richness values, consistent with the scatter
observed in the X-ray luminosity-richness relation studied in
\citet{Rykoff+12}. We are currently extending the current sample to groups with a lower number
of members \citep{erfanianfar+14}.

To extend our group sample to lower redshifts, we complement it with
the stacked groups of \citet{Guo+14}. \citet{Guo+14} stack the
optically selected group sample of \citet{Robotham+11} drawn from the
GAMA survey over an area of 135 $\rm{deg}^2$ to derive the group
$L_{\nu}(250 {\mu}m)$ LF in the local Universe. The groups are stacked
in several redshift and halo mass bins, from $z=0$ to 0.4 in bins of
0.1, and from $10^{12} M_{\odot}$ to $10^{14} M_{\odot}$ in bins of
0.5 dex.  In particular, we include in our sample the stacked groups
with masses above $10^{12.5} M_{\odot}$, for consistency with our halo
mass cut, and at $z < 0.2$, as at higher redshift the IR LF of the
stacked groups is only poorly constrained, given that only the very
high luminosity end is observed.  To extend the group sample to higher
redshifts, up to $z\sim 1.6$, we include the GOODS-S group
identified by \citet{Kurk+08} and the one studied by \citet{Smail+14}
at $z \sim 1.6$. Both groups are covered by deep PACS or SPIRE
observations. The former structure was initially optically detected
through the presence of an overdensity of [OII] line emitters by
\citet{vanzella+06} and, then, as an overdensity of elliptical
galaxies by \citet{Kurk+08} in the GMASS survey, and is also an X-ray
group candidate, as found in \citet{tanaka+13} in the CDFS (see also
Popesso et al. 2014).  The latter structure, Cl $0218.3-0510$, lies in
the UKIRT Infrared Deep Sky Survey/Ultra-Deep Survey field of the
SCUBA-2 Cosmology Legacy Survey.

To extend the dynamical range studied in this work, we include also
the individual clusters studied in \citet{popesso+12} and observed with
PACS in the PEP survey \citep{Lutz+11} at $0 < z < 1$, with the
exclusion of the Bullet cluster, which is a peculiar system with an
ongoing merging process of a cluster and a group. We include also the
Coma cluster of Bai et al. (2006) observed with {\it{Spitzer}} MIPS
and the three stacked clusters derived by \citet{Haines+13} from a
sample of 33 LoCuSS clusters observed with PACS and SPIRE at $0.15 <z
< 0.2$, $0.2 < z < 0.25$ and $0.25 < z <0.3$.

We take as estimate of the total mass of the system ($M_{halo}$) the
mass $M_{200}$ enclosed within a sphere of radius $r_{200}$, where
$r_{200}$ is the radius where the mean mass overdensity of the
group/cluster is 200 times the critical density of the Universe at the
group mean redshift. For the group sample of Popesso et al. (2014),
the total mass of the groups is derived from their X-ray luminosity
($L_X$) by using the $L_X-M_{200}$ relation of
\citet{Leauthaud+10}. The total mass $M_{200}$ of the stacked group of
\citet{Guo+14} is given by the mean mass of the corresponding halo
mass bin. The mean mass of the stacked LoCuSS clusters is given by
\citet{Haines+13} as the mean of the cluster masses contributing to
each stack. The mass of the Coma cluster and Cl $0218.3-0510$ are taken
from Bai et al. (2006) and Smail et al. (2014), respectively.

\begin{figure}
\begin{center}
\includegraphics[width=0.49\textwidth,angle=0]{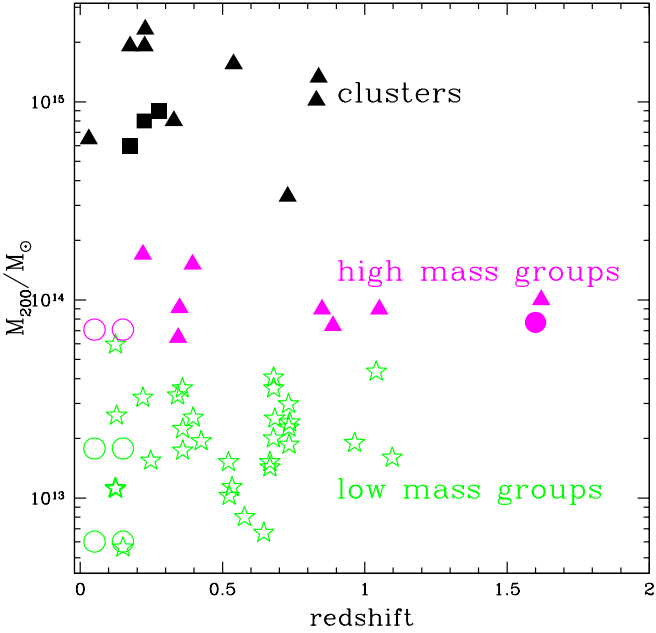} 
\caption{Total mass ($M_{200}$) versus redshift of the group and
  cluster samples. The low mass (poor) groups ($6{\times}10^{12}<
  M_{200} < 6{\times}10^{13}$ $M_{\odot}$) are identified by the green
  symbols. The high mass (rich) groups ($6{\times}10^{13} < M_{200} <
  2{\times}10^{14}$ $M_{\odot}$) are identified by the magenta
  symbols. Black symbols show the sample of clusters, all with
  $M_{200} > 2{\times}10^{14}$ $M_{\odot}$. The stacked groups of Guo
  et al. (2014) are shown with a empty circles. The stacked clusters
  of Haines et al. (2013) are shown with filled squares. The high
  redshift group of Smail et al. (2014) is shown with a filled
  point. Green stars and filled magenta triangles show, respectively,
  the low mass and high mass groups of Popesso et al. (2014). Filled
  black triangles show the cluster sample of Popesso et al. (2012)
  with the addition of the Coma cluster of Bai et al. (2006). The typical error of $M_{200}$ 
for the low and high mass group samples is of the order of 0.2 dex. 
Thus, much smaller than the $M_{200}$ bin size considered in this analysis. 
The typical error of the bright clusters is of the order of 10\% in the totality of the cases.}
\label{mass_redshift}
\end{center}
\end{figure}

The group and cluster masses ($M_{200}$) vs. redshifts are shown in
Fig. \ref{mass_redshift}. We collect in total a sample of 57
systems. For this particular analysis we also distinguish among low
mass and high mass groups. Galaxy groups with masses in the range
$6{\times}10^{12}-6{\times}10^{13}$ $M_{\odot}$ are considered
low mass systems. Galaxy groups with masses in the range
$6{\times}10^{13}-2{\times}10^{14}$ $M_{\odot}$ are considered
high mass systems. All clusters used in this analysis have masses
above $2{\times}10^{14}$ $M_{\odot}$.

\subsection{Bolometric IR luminosity}
\label{s:sfr}
For all the groups in the sample of Popesso et al. (2014) is the
membership of their galaxies available.  For the galaxy members
observed either by {\it{Herschel}} PACS or by {\it{Spitzer}} MIPS, we
compute the IR luminosities integrating the spectral energy
distribution (SED) templates from \citet{Elbaz+11} in the range
8-1000\,$\mu$m. The PACS (70, 100 and 160\,$\mu$m) fluxes, when
available, together with the 24 $\mu$m fluxes are used to find the
best fit templates among the main sequence (MS) and SB
\citep{Elbaz+11} templates. When only the 24 $\mu$m flux is available
for undetected PACS sources, we rely only on this single point and we
use the MS template for extrapolating the $L_{IR}$. Indeed, the MS
template turns out to be the best fit template in the majority of the
cases (80\%) with common PACS and 24 $\mu$m detection \citep[see][for
  a more detailed discussion]{ziparo+13}. In principle, the use of the
MS template could cause only an under-estimation of the extrapolated
$L_{IR}$ from 24~$\mu$m fluxes, in particular at high redshift or for
off-sequence sources due to the higher PAHs emission of the MS
template \citep{Elbaz+11,nordon+10}. However, as shown in
\citet{ziparo+13}, the comparison between the $L_{IR}$ estimated with
the best fit templates based on PACS and 24 $\mu$m data, and the
$L_{IR}$ extrapolated from 24 $\mu$m flux only with the MS template
($L_{IR}^{24} $), shows the two estimates are in very good agreement,
with a slight discrepancy (10\%) only at $z\geq 1.7$ or at
$L_{IR}^{24} > 10^{11.7}\ L_\odot$.

\begin{figure}
\begin{center}
\includegraphics[width=0.49\textwidth,angle=0]{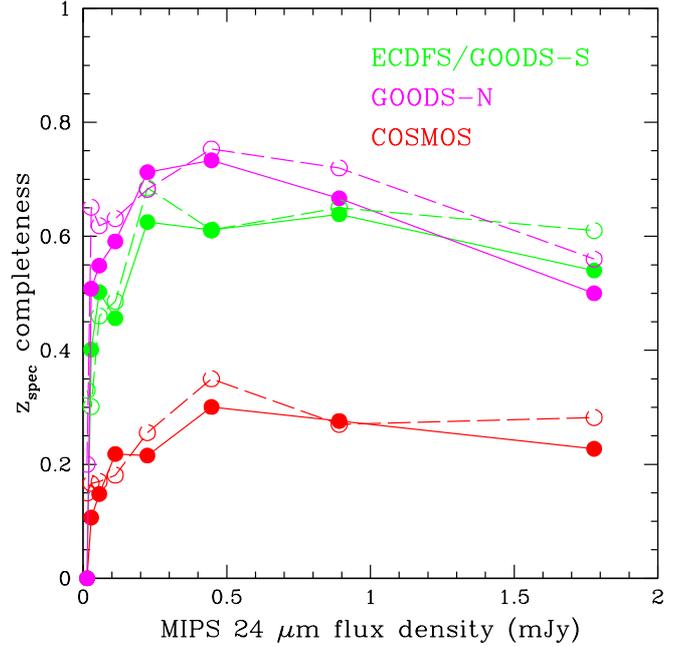} 
\caption{Mean spectroscopic completeness in the {\it{Spitzer}} MIPS 24
  $\mu$m band across the whole areas of the ECDFS, GOODS-N and COSMOS
  field (solid lines) and mean spectroscopic completeness simulated
  in the ``incomplete'' mock catalogs (dashed lines).}
\label{histogram}
\end{center}
\end{figure}

\section{Estimate of the  total SFR and SFR/M}
\label{estimate}

In this work we define the total star formation rate of a galaxy
system (\tsfra) as the sum of the SFR of its galaxy members with
$L_{IR}$ down to $10^7$ $L_{\odot}$. The total SFR per
unit of halo mass is defined as the \tsfra~ divided by the total
mass of the galaxy system. We explain here how these quantities are
estimated for the different halo subsamples.

For the Popesso et al. (2012, 2014) group and cluster samples, the IR
emitting -- spectroscopically identified -- galaxy members are
available. Thus, for the groups and clusters of this subsample, the
total IR luminosity of each system is obtained by summing up the
$L_{IR}$ of the members within the system $r_{200}$ and down to the
$L_{IR}$ limit ($L_{IR,limit}$) corresponding to the $5{\sigma}$ flux
level ($f_{limit}$) of the deepest IR band, which is {\it{Spitzer}}
MIPS 24 $\mu$m for all the groups of Popesso et al. (2014) and
{\it{Herschel}} PACS 100 $\mu$m for several clusters of Popesso et
al. (2012, see also Popesso et al. 2014 for the details about the flux
limits reached in different fields and in different bands).  The total
IR luminosity is then converted into a total SFR via the \citet{Kennicutt98} relation.
In this conversion we assume that the IR flux is
completely dominated by obscured SF and not by AGN activity also for
the 5\% AGNs identified as X-ray sources among the group galaxy
members. 87\% of these AGNs are bright IR emitting galaxies observed by
PACS. \textit{Herschel} studies of X-ray AGNs
\citep{Shao+10,mullaney+12,rosario+12} have demonstrated that in the
vast majority of cases (i.e., $>\,$94\%) the PACS flux densities are
dominated by emission from the host galaxy and thus provide an
uncontaminated view of their star-formation activities. The flux at 24
${\mu}$m of the remaining 13\% of AGNs observed only by {\it{Spitzer}}
could in principle be contaminated by the AGN emission. However, since
these galaxies are faint IR sources and they represent only the 0.65\%
of the group galaxy population studied in this work, we consider that
their marginal contribution can not affect our results. Consequently,
we assume that the IR luminosity derived here has no significant
contribution from AGNs and can be converted into the obscured total
SFR density of group galaxy population.

We also correct \tsfra~ for spectroscopic incompleteness by
multiplying it by the ratio of the number of sources without and with
spectroscopic redshift, with flux density larger than $f_{limit}$ and
within $3{\times}r_{200}$ from the X-ray center of the system. The
incompleteness correction is estimated within $3{\times}r_{200}$
rather than within $r_{200}$ to increase the statistics and to have a
more reliable estimate. This correction is based on the assumption
that the spectroscopic selection function is not biased against or in
favor of group or cluster galaxies, as ensured by the very
    homogeneous spatial sampling of the various spectroscopic campaigns
    conducted in the considered fields (see Cooper et al. 2012 for
    ECDFS, Barger et al. 2008 for GOODS-N and Lilly et al. 2009 for
    COSMOS). The incompleteness correction factor ranges from 1.2 to
1.66.  To extend \tsfra~ down to $L_{IR}=10^7$ $L_{\odot}$, we use the
group and cluster IR Luminosity Function (LF). This is estimated for
the groups in several redshift bin up to $z\sim 1.6$ in Popesso et
al. (2014). For low redshift clusters it is provided by
\citet{Haines+10}, for the intermediate redshift LoCuSS clusters in
Haines et al. (2013), and for the high redshift clusters at $0.6 < z <
0.8$ by \citet{Finn+10}. For clusters outside the mentioned redshift
bins, we use the IR LF of the closest redshift bin. The best fit LFs
are estimated in a homogeneous way for all cases in Popesso et
al. (2014). We use here the best fits obtained with the modified
Schechter function of \citet{saunders+90}. The correction down to
$L_{IR}=10^7$ $L_{\odot}$ is estimated as the ratio between the
integral of the group or cluster IR LF down to $L_{IR}=10^7$
$L_{\odot}$ and the integral down to the $L_{IR,limit}$ of each
system. The correction down to $L_{IR} =10^{7} L_{\odot}$ due to the
extrapolation from the best fit IR LF of groups and clusters is
limited to less than 10-20\% in all cases. Indeed, for low redshift
systems, \tsfra~ is estimated down to very faint $L_{IR}$ and a very
small correction is applied. At higher redshift, instead, as shown in
Popesso et al. (2014) for the groups and in Haines et al. (2013) for
the clusters, the bulk of the IR luminosity is provided by the LF
bright end and a marginal contribution is provided by galaxies at
$L_{IR} < 10^{10} L_{\odot}$. Thus, even for high redshift groups and
clusters, whose \tsfra~ estimate is limited to the IR brightest
members, the correction down to $L_{IR}=10^7$ $L_{\odot}$ is small.

To test the reliability of our method, in particular of the
spectroscopic incompleteness correction, we use the mock catalogs of
the Millennium Simulation \citep{springel+05}. Out of several mock
catalogs created from the Millennium Simulation, we choose those of
\cite{KW07} based on the semi-analytical model of
\cite{delucia+06}. Kitzbichler \& White (2007) make mock observations
of the artificial Universe by positioning a virtual observer at
$z\sim0$ and finding the galaxies which lie on his backward
light-cone. We select several mock light-cones catalog and extract from
those the following info for each galaxy: the Friend of Friend (FoF)
identification number, to identify the galaxy member of the same
group/cluster (same FoF ID), the dark matter halo virial mass which,
according to De Lucia et al. (2006), is consistent with the mass
calculated within $r_{200}$, as in the observed sample, the SFR and
the redshift. We transform the SFR into $L_{IR}$ by following the
\citet{Kennicutt98} relation. We use, then, the MS template of
\citet{Elbaz+11}, redshifted to the galaxy redshift, to estimate the
      {\it{Spitzer}} MIPS 24 $\mu$m flux of each simulated galaxy.
      The SB template is not used as off-sequence galaxies are
      generally much less numerous than MS ones.

In order to simulate the effects of the spectroscopic selection
function of the surveys used in this work, we randomly extract as a
function of the simulated MIPS 24 $\mu$m flux bin, a fraction of
galaxies consistent with that of galaxies with available spectroscopic
redshift in the same flux bin, observed in the GOODS and COSMOS
surveys (see Fig. \ref{histogram}).  Since the clusters of the
\citet{popesso+12} subsample are characterized by a spectroscopic
completeness intermediate between the GOODS and COSMOS surveys, we
limit this analysis to these surveys.  We randomly extract 25 catalogs
for each survey from different light-cones. The ``incomplete'' mock
catalogs, produced in this way, tend to reproduce, to a level that
we consider sufficient for our needs, the biased selection towards
highly star forming galaxies observed in the real galaxy samples (see
dashed lines in Fig. \ref{histogram}).

\begin{figure}
\includegraphics[width=0.45\textwidth,angle=0]{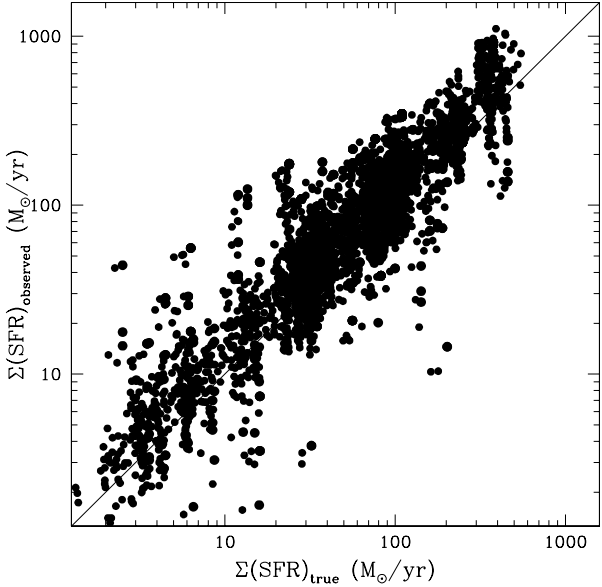}
\caption{``observed'' vs. ``true'' values of total SFR in the mock
  catalogs of Kitzbichler \& White (2007)}
\label{incomplete}
\end{figure}

We extract, then, from the original Kitzbichler \& White (2007) mock
catalogs a sample of galaxy groups and clusters in the same mass and
redshift range as those of the observed sample. The members of the
structures are identified by the same FoF identification number,
defined according to the FoF algorithm described in De Lucia et
al. (2006). We estimate the ``true'' \tsfra~ as the one obtained by
summing up the SFR of all members within $r_{200}$. We limit this
estimate down to $10^{10}$ $L_{\odot}$ for all the structures as we
can not correct down to $10^7$ $L_{\odot}$ since the IR LF of groups
in the mock catalog is not known and its determination is outside the
scope of this paper. We point out that $10^{10}$ $L_{\odot}$ is the
average $L_{IR,limit}$ reached in our data-set. We estimate the
``observed'' \tsfra~ by summing up the member SFR of the same
structures in the incomplete catalogs and by correcting for
incompleteness by following the same procedure applied in the real
data-set.  Figure \ref{incomplete} shows the comparison of the ``true''
and "observed" quantities. We find a rather good agreement between the
two values with a scatter about 0.2 dex. We use these simulations to
estimate the error due to incompleteness in the \tsfra~. This is
estimated as the dispersion of the distributions of the residual
${\Delta}(\rm{SFR})={\Sigma}(\rm{SFR})_{\rm{true}}-{\Sigma}(\rm{SFR})_{\rm{observed}}$. This
uncertainty varies as a function of the completeness level and of the
number of group members. For a given completeness level, the lower the
number of group members, the higher the uncertainty.

The total uncertainty of the \tsfra~ estimates is determined from the
propagation of error analysis, by considering a 10\% uncertainty in
the \lir estimates \citep[see][for further details]{Lutz+11}, and the
uncertainty due to the completeness correction.  We do not consider
the error of the correction down to $10^7$ $L_{\odot}$ since this
correction is marginal.

For the Guo et al. (2014) stacked groups, we estimate in each redshift
and halo mass bin, respectively, the total IR luminosity by
integrating the corresponding IR LF. As explained in Popesso et
al. (2014), Guo et al. (2014) provide the {\it{Herschel}} SPIRE 250
$\mu$m LF, which must be converted into a total IR LF. For this
purpose we use equation no. 2 of Guo et al. (2014) to transform the
group $L_{\nu}(250 {\mu}m)$ LF into the group IR LF. For consistency
with Popesso et al. (2014), we fit the total IR LF of the Guo et
al. (2014) stacked groups with the modified Schechter function of
\citet{saunders+90}. Due to the very low statistics of the Guo et
al. (2014) LF in the $0.3 < z <0.4$ redshift bin, we limit this
analysis to the $z < 0.3$ groups. The total IR luminosity of each
stacked group is obtained by integrating the best fit modified
Schechter function down to $10^7$ $L_{\odot}$. This is, then,
converted into the \tsfra~ via the \citet{Kennicutt+98} relation. The
error in \tsfra~ is obtained by propagating the error of the total IR
luminosity, which is in turn obtained by marginalizing over the errors
of the best fit parameters.

For the stacked LoCuSS clusters of Haines et al. (2013) we use a
slightly different approach. For the three stacked clusters at $0.15<
z < 0.2$, $0.2, z < 0.25$ and $0.25 < z <0.3$ Haines et al. (2013)
provide the \tsfra~ obtained by integrating the cluster IR LF down to
$10^{11}$ $L_{\odot}$. This was done to compare the LoCuSS cluster
\tsfra~ with the results of Popesso et al. (2012), where the \tsfra~
estimate was limited to the LIRG population. We use the best fit
of the IR LF of Haines et al. (2013) obtained by Popesso et al. (2014) 
to correct this quantity down to $10^7$ $L_{\odot}$. The error in the
\tsfra~ is obtained by summing in quadrature the errors of the
estimates provided by Haines et al. (2013) and the error of the
correction which is obtained by marginalizing over the errors of the
best fit parameters. The \tsfra~ of the Coma cluster is obtained by
integrating the IR LF of Bai et al. (2006) down to $10^7$
$L_{\odot}$. Also in this case the error is obtained by marginalizing
over the errors of the best fit parameters.

The IR LF of the Kurk et al. (2008) structure is studied in Popesso et
al. (2014). We integrate this LF down to $10^7$ $L_{\odot}$ to obtain
the total IR luminosity of the structure and, thus, its \tsfra~ via
the \citet{Kennicutt98} relation. Also in this case the error is
estimated by marginalizing over the errors of the best fit
parameters. For Cl $0218.3-0510$, Smail et al. (2014) provide the
\tsfra~ obtained by summing up the contribution of the LIRGs in the
structure. Also in this case this was done to compare with the results
of Popesso et al. (2012). We use the IR LF of the Kurk et al. (2008)
structure at the same redshift, to correct the estimate of Smail et
al. (2014) for the contribution of galaxies with
IR luminosity in the $10^7-10^{11}$ $L_{\odot}$ range. The error in
the \tsfra~ is obtained by summing in quadrature the errors of the
estimates provided by Smail et al. (2014) and the error of the
correction.

We finally define the total SFR per unit halo mass ($\tsfrma$) as the
ratio of \tsfra~ and the dynamical mass of the system within
$r_{200}$, $M_{200}$. The error is estimated by propagating the error
on \tsfra~ and the error on the mass. By taking into account also the
error due to the incompleteness correction applied to the
\tsfr, we obtain that the accuracy of the $\tsfrma$ estimate
is $\sim$0.25-0.3 dex.

\begin{figure*}
\begin{center}
\includegraphics[width=0.49\textwidth]{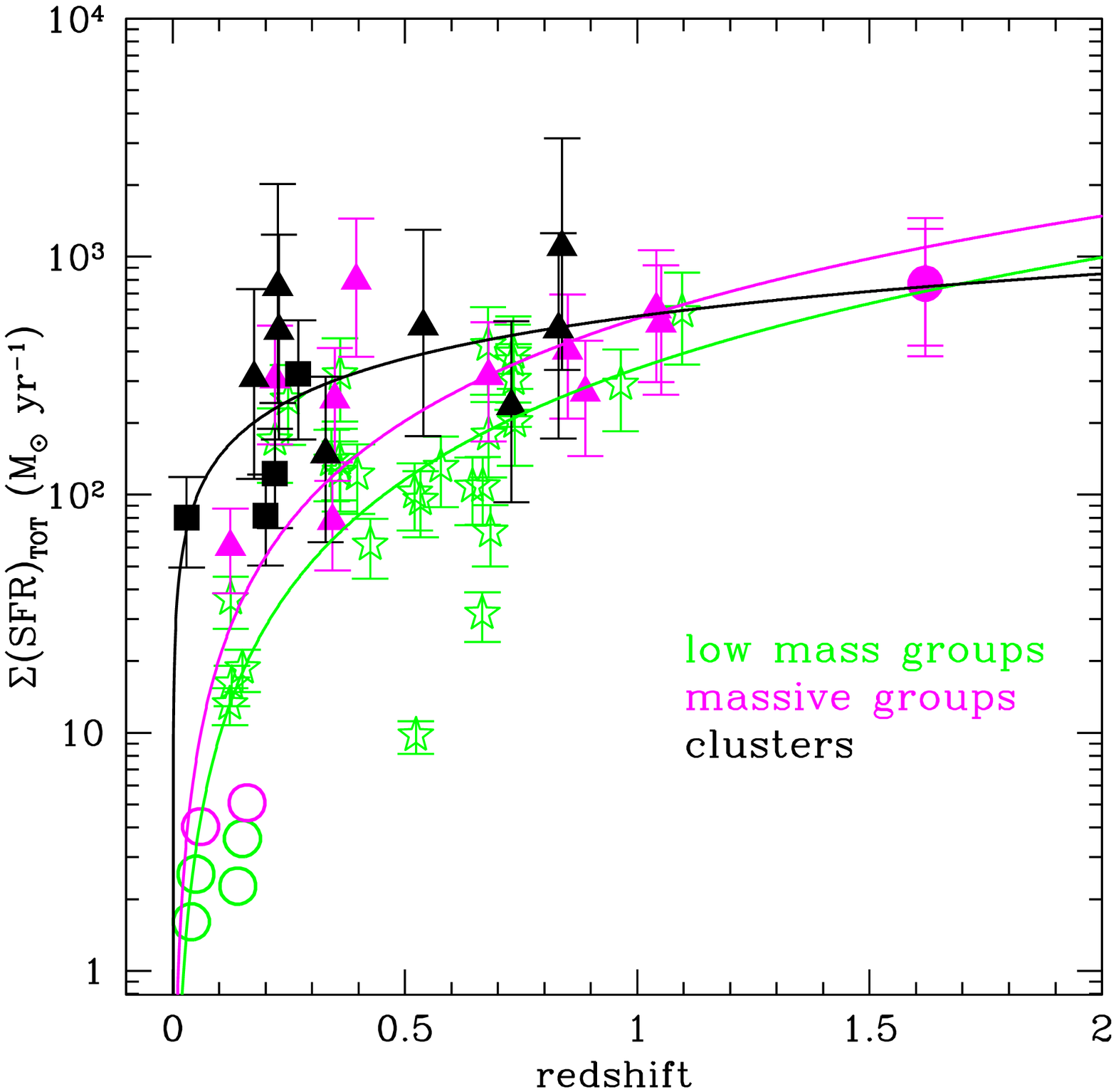}
\includegraphics[width=0.49\textwidth]{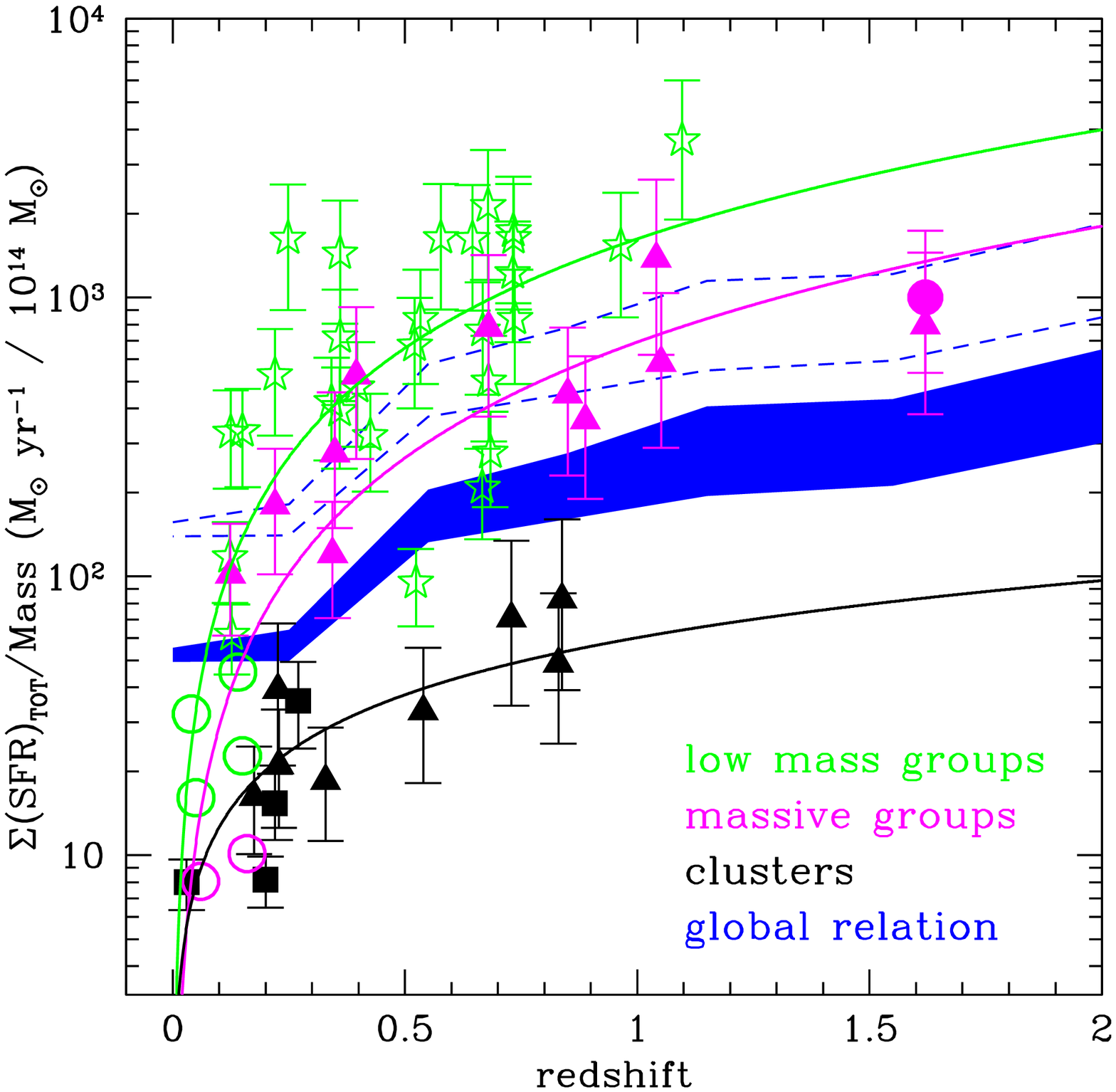}
\caption{{\it{Left panel}}: $\Sigma$(SFR)- redshift relation for low
  mass groups (green symbols), massive groups (magenta symbols) and
  clusters (black symbols) up to redshift $\sim$ 1.6. The group and
  cluster sample of Popesso et al. (2012, 2014) are shown with
  triangles. The stacked groups of Guo et al. (2014) are shown with
  empty circles. The stacked clusters of Haines et al. (2013) and the
  Coma cluster of \citet{Bai+06} are indicated with filled
  squares. The Kurk et al. (2008) group is shown with a triangle while
  the Smail et al. (2014) structure is shown with a filled circle. The
  green, magenta and black solid lines show the low mass group,
  massive group and cluster best fit relation of the form $\Sigma$(SFR)
  $\propto z^{\alpha}$, respectively. {\it{Right panel}}: $\tsfrma$-
  redshift relation for the same sample. The color coding and the
  symbols have the same meaning as in the left panel. The global
  $\tsfrma$-redshift relation is derived from \citet{magnelli+13} and
  it is shown by the blue shaded region. The shading represents
  1$\sigma$ confidence levels as reported in \citet{magnelli+13}. This
  region is moved $0.4$ dex up (two dashed blue lines) to indicate
  its locus under the assumption that not all the mass in the
  considered volume is locked in halos, as indicated by
  \citet{Faltenbacher+10}.}
\label{fig_1}
\end{center}
\end{figure*}

\section{The total SFR per halo mass versus redshift}
\label{sfr}

In Fig. \ref{fig_1} we show the \tsfra-$z$ (left panel) and the
$\tsfrma-z$ (right panel) relations for the systems considered in this
work. As explained in Section \ref{s:data}, we distinguish the galaxy
systems in low mass (green points) and high mass (magenta points)
groups, and clusters (black points), depending on their total mass. We
also show the global relation. This global relation (blue shaded
region in the figure) is obtained by dividing the observed Star
Formation Rate Density (SFRD) of \citet{magnelli+13}, by the mean
comoving density of the universe ($\Omega_m \times \rho_c $ where
$\Omega_m=0.3$ and $\rho_c$ is the critical density of the
Universe). The SFRD has been evaluated by integrating the global IR LF
of \citet{magnelli+13} based on PACS data, down to
$L_{IR}=10^7/L_{\odot}$ and by converting the integrated IR luminosity
into a total SFR via the \citet{Kennicutt98} relation in each redshift
slice. The global SFRD has been estimated in large comoving volumes
that include galaxy systems, voids, and isolated galaxies, and is thus
representative of the global galaxy population. We point out, however,
that the $\tsfrm$ for the global population is only a lower
limit. Indeed, not all the mass is locked in halos hosting
galaxies. As shown in \citet{Faltenbacher+10} the fraction of mass
locked in halos also depends on the local density field. Thus, using
the mean comoving density of the universe can lead to under-estimate
the $\tsfrm$ for the global galaxy population. Following
\citet{Faltenbacher+10} this under-estimation should be $\leq 0.4$
dex. In the right panel of Fig. \ref{fig_1} we move the global
relation upwards of $0.4$ dex to show where it should lie under the
assumption that not all the mass is locked in dark matter halos.

\begin{figure*}
\begin{center}
\includegraphics[width=0.49\textwidth]{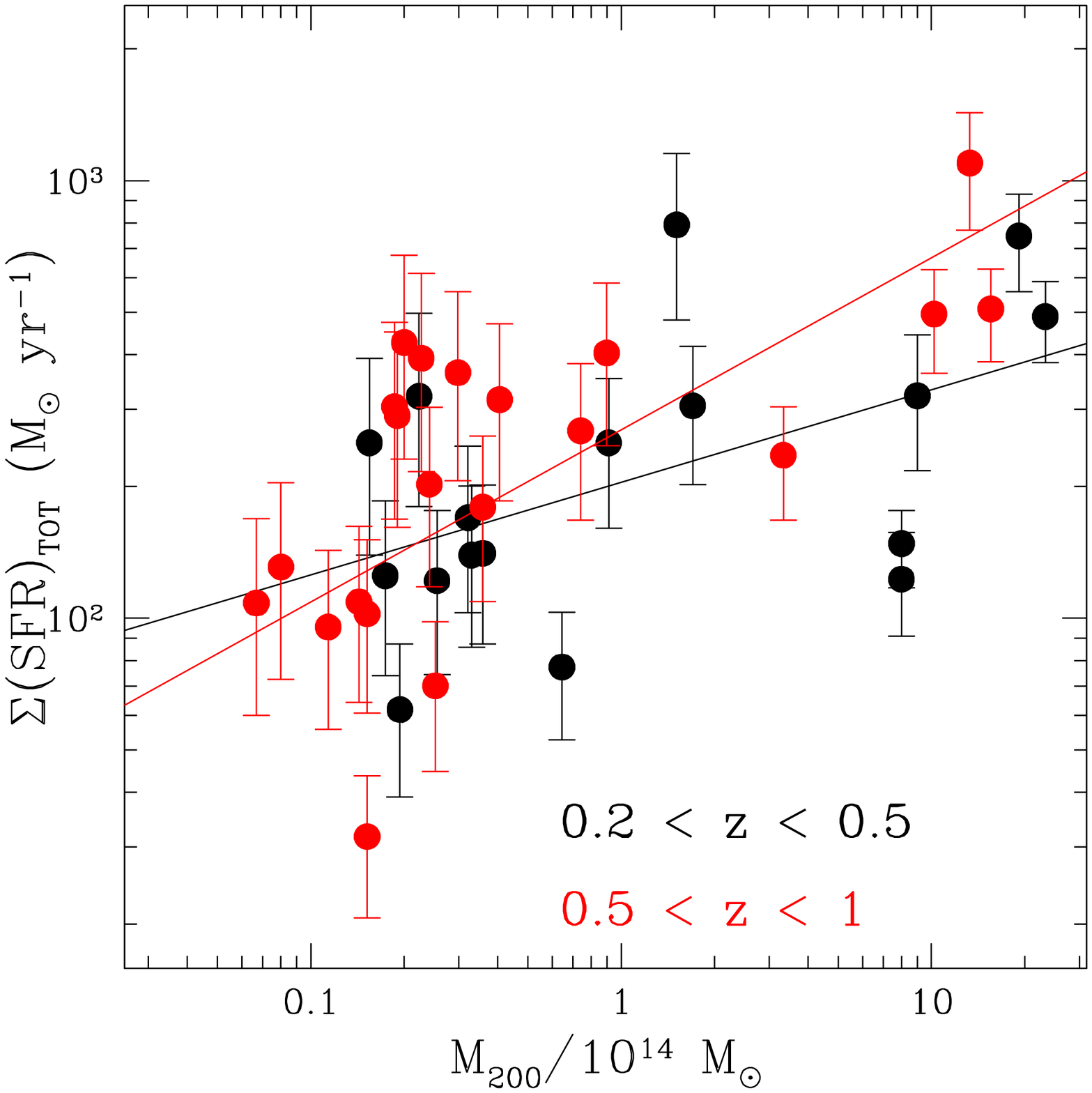} 
\includegraphics[width=0.49\textwidth]{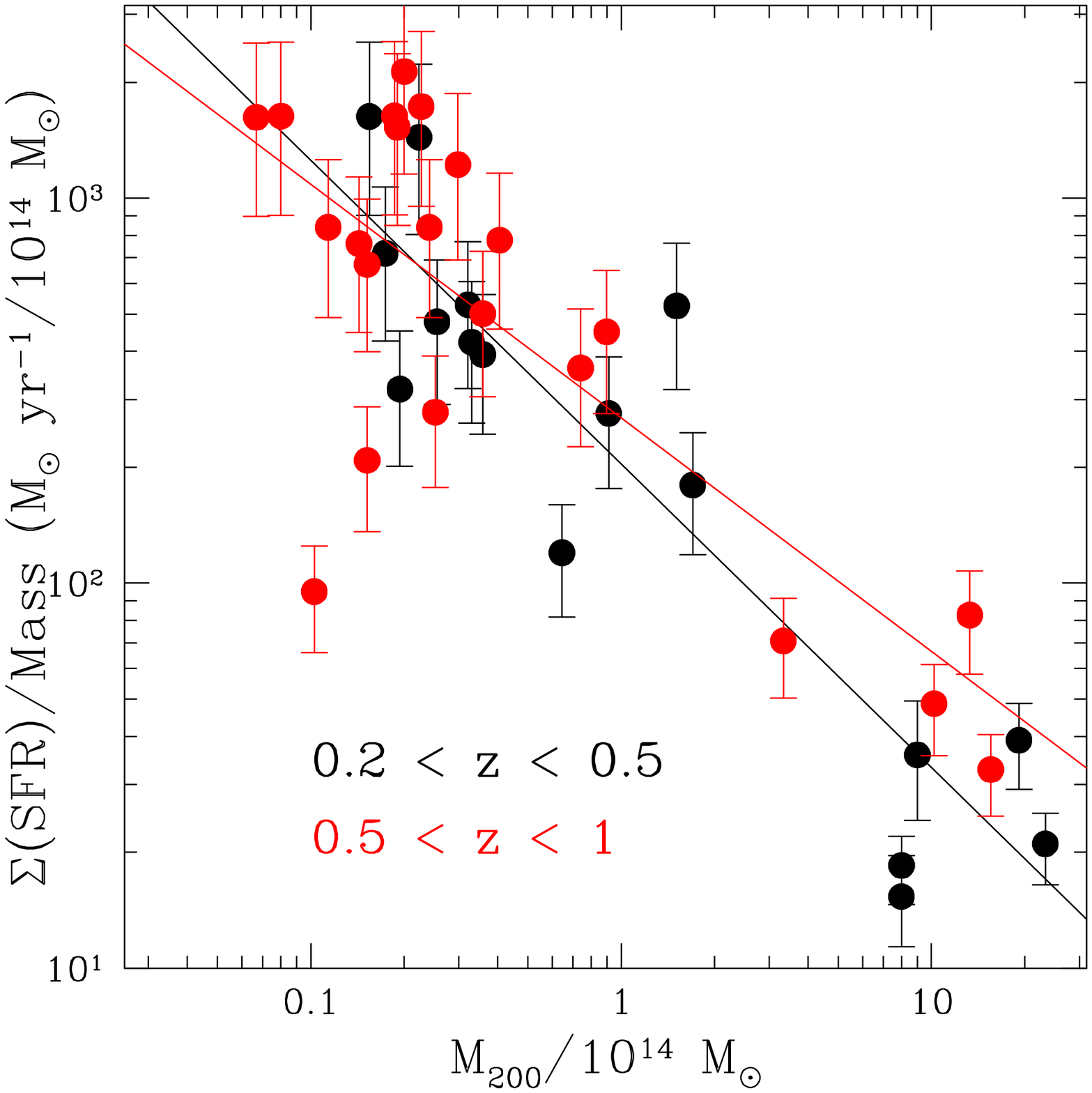} 
\caption{ $\Sigma$(SFR)--mass relation in two redshift slices, $0.2 <
  z < 0.5$ (red points) and $0.5<z <1$ (black points) of
  Fig. \ref{fig_1}. The solid lines show the best fit
  relations. $\tsfrma$--mass relation in two redshift slices, $0.2 < z
  < 0.5$ (red points) and $0.5<z <1$ (black points) of
  Fig. \ref{fig_1}. The solid lines show the best fit relations.}
\label{zoom}
\end{center}
\end{figure*}

We fit both the \tsfra$-z$ and the $\tsfrma-z$ relations with a power
law. The best-fits in the three mass bins for the $\Sigma(SFR)-z$
relation are: for clusters

\begin{equation}
$\tsfra$=(562\pm 38) \times z^{0.6\pm0.2}
\end{equation}
\begin{equation}
\tsfrma=(60\pm 18) \times z^{0.7\pm0.2} 
\end{equation}
for massive groups:
\begin{equation}
$\tsfra$=(550\pm 45) \times z^{1.4\pm0.3}
\end{equation}
\begin{equation}
\tsfrma=(691\pm 45) \times z^{1.4\pm0.2} 
\end{equation}
and for low mass groups:
\begin{equation}
$\tsfra$=(338\pm 43) \times z^{1.6\pm0.2} 
\end{equation}
\begin{equation}
\tsfrma=(1617\pm 223) \times z^{1.3\pm0.2} 
\end{equation}

The \tsfra$-z$ is much noisier than the $\tsfrma-z$ relation. The
power law fit is poorly constrained in the \tsfra$-z$ relation while it
provides a very good fit for the $\tsfrma-z$ relation. The total SFR
appears to be rather similar in all structures independently of the
redshift. Clusters are in general much richer in number of galaxies
with respect to low mass groups, thus, their total SFR is on average
higher (0.2-0.3 dex, $1-1.5 \sigma$) with respect to the low mass
systems. Indeed, if we zoom into a redshift slice, as shown for
instance in the left panel of Fig. \ref{zoom}, we observe a rather
flat, though clear, positive correlation between the total SFR and the
system mass (\tsfr $ \propto M^{0.2}$ at $0.2 < z < 0.5$ and
\tsfr $ \propto M^{0.35}$ at $0.5 < z < 1$). Since the number
of galaxies is increasing linearly with the halo mass, as shown by
\citet{Yang+07} in the local Universe, and more recently by Erfanianfar
et al. (in prep) up to $z \sim 1$, it follows that at least up to $z
\sim 1$ the mean SFR is higher in the low mass systems than in
clusters.

Once the $\Sigma(SFR)$ is normalized to the halo mass, the situation
reverses, and the low mass groups appear to be more active per unit
mass than their high mass counterparts. Indeed, as shown in the right
panel of Fig. \ref{fig_1}, the low mass groups show a mean activity
per unit mass more than one order of magnitude higher with respect to
the clusters. The figure, thus, indicates a clear anti-correlation
between $\tsfrma$ and the system mass at any redshift. The $\tsfrma-
M_{200}$ relation exhibits a higher significance with respect to the
$\Sigma$(SFR)$- M_{200}$ relation, and a lower scatter, as shown in
the right panel of Fig. \ref{zoom}. We point out that we do not find a
faster evolution (steeper relation) in more massive systems, unlike
\citet{popesso+12}. However, at variance with \citet{popesso+12} we
consider here the global IR emitting population by integrating the
system IR LF while \citet{popesso+12} consider only the evolution of
the LIRG population. In Popesso et al. (2014) we show that the LIRG
population is evolving in a much faster way in massive systems and
this could be the cause of the apparent inconsistency with our
previous results. In addition, we here also split groups into two
subsamples, and we have to pay the price of a poorer statistics and,
thus, larger errors in the best fit parameters.

\begin{figure}
\begin{center}
\includegraphics[width=0.49\textwidth]{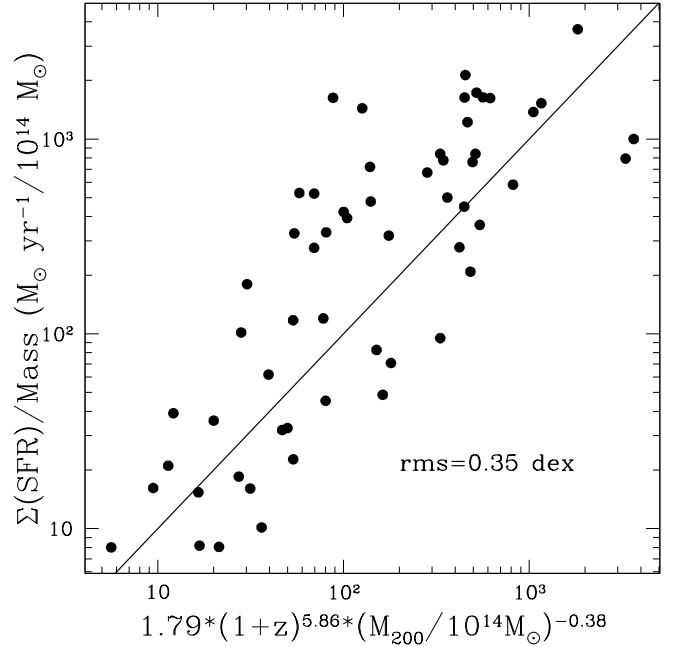} 
\caption{Residual of the observed $\tsfrma$ with respect to the best fit plane $\tsfrma-M_{200}-z$.}
\label{residual}
\end{center}
\end{figure}

Lower mass groups appear to lie above the global relation, something
that was not noted in our previous analysis \citep{popesso+12}.  A lot
of star formation activity is therefore occurring in the small volume
occupied by the numerous group-sized dark matter halos.  More massive
groups tend to have a SF activity per halo mass consistent with the
global relation, even if we account for an under-estimation of the
global $\tsfrm$ of 0.4 dex \citep{Faltenbacher+10}.  Star formation
activity is largely suppressed in the most massive, cluster-size,
halos at any redshift.

In order to take full advantage of the redshift and dynamical range
covered by our sample, we also fit the the $\tsfrma-M_{200}-z$
plane. The best fit turns out to be of the form:
\begin{equation}
\tsfrma=(1.8\pm 0.3)\times (1+z)^{5.9\pm0.8}\times M_{200}^{-0.38\pm0.06}
\end{equation}

The scatter around the plane is 0.35 dex, which is slightly larger
than the accuracy in our estimate of the $\tsfrma$, which, according
to our simulation (see Section \ref{estimate}), is of 0.25-0.3 dex. We
point out that, while for the $\tsfrm-z$ relation in the individual
mass bin, a power law of the form $\tsfrm \propto z^{\alpha}$ provides
the best fit in all cases, a redshift dependence of the kind $\tsfrm
\propto (1+z)^{\alpha}M_{200}^{\beta}$, provides a slightly better fit
to the plane. Indeed the final scatter around the plane decrease from
0.42 dex to 0.35 dex as shown in Fig. \ref{residual}. We try to adopt
the same approach also for the $\Sigma$(SFR)$-z-M_{200}$ plane, but
the large scatter observed in the $\Sigma$(SFR)$-z$ relation generates
a large scatter also in the plane.

\begin{figure}
\begin{center}
\includegraphics[width=0.49\textwidth]{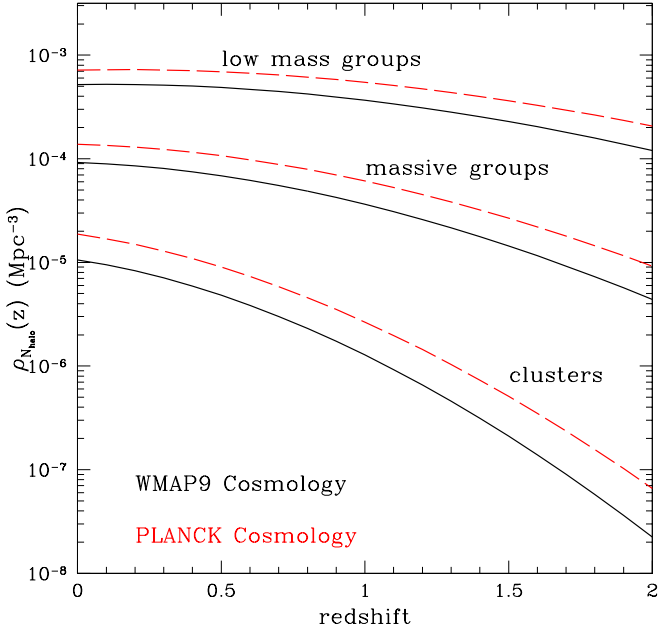} 
\caption{Comoving density of the dark matter halos as a function of
  redshift ($\rho_{N_{halo}}(z)$) in the three $M_{200}$ mass ranges
  considered in this work according to the WMAP9 (black solid curves)
  and Planck (red dashed curves) cosmologies.}
\label{g_density}
\end{center}
\end{figure}

\section{The Cosmic Star Formation Rate Density of massive halos}
\label{s:sfr_density} 

\begin{figure}
\begin{center}
\includegraphics[width=0.49\textwidth]{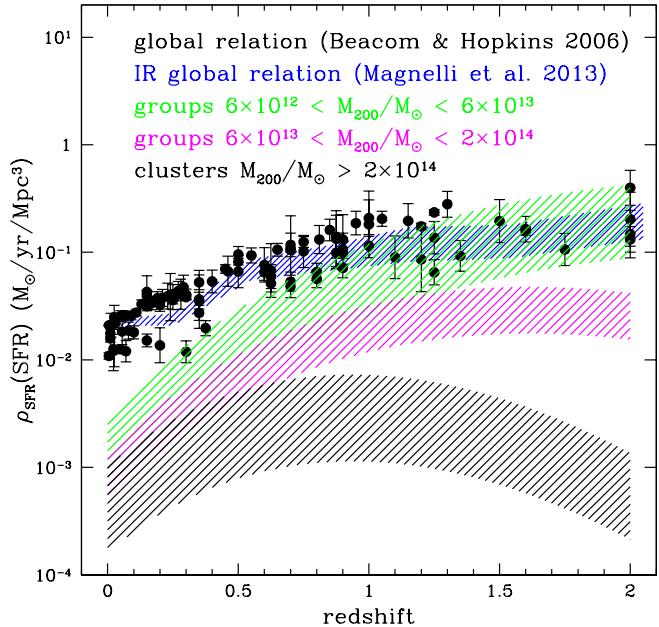}
\caption{Contribution of each halo mass range to the CSFH as a
  function of redshift. The CSFH of the Universe is taken from
  \citet{magnelli+13}. The shading indicates the 1$\sigma$ confidence
  level as derived in \citet{magnelli+13}.  The shaded green, magenta,
  and black regions show the contributions of the low mass group,
  high mass group, and cluster galaxy populations, respectively. The
  black points show the compilation of \citet{Hopkins_Beacom06}.}
\label{sfr_density}
\end{center}
\end{figure}

To understand what is the contribution of DM halos of different masses
to the evolution of the Cosmic SFRD ($\sfrd$), we use the best fit
$\tsfrm-z-M_{200}$ plane with the following procedure. Since \tsfr is
a monotonically increasing function of $M_{200}$ (see left panel of
Fig. \ref{zoom}), we use the lower and upper limit of each mass range
to retrieve the corresponding lower and upper limit of the $\Sigma$(SFR)$-z$
relation by fixing the value of $M_{200}$ in the $\tsfrm-z-M_{200}$
plane.  We prefer this approach rather than using directly the fitted
\tsfr$-z-M_{200}$ plane and the \tsfr$-z$ relation because, as
discussed in previous section, the larger noise of these correlations
leads to a poorer fit in comparison to the $\tsfrm-z-M_{200}$ plane.
To transform the $\Sigma$(SFR)$-z$ regions identified for each mass bin
into a SFR density as a function of redshift, we multiply each of them
for the comoving number density of DM halos in the corresponding mass
range as a function of redshift ($\rho_{N_{halo}}(z)$). This quantity
is estimated by using the WMAP9 concordance model prediction of the
comoving $\rho_{N_{halo}}(z)$ of halos in the three mass ranges. This model
reproduces the observed $\log(N)-\log(S)$ distribution of the deepest
X-ray group and cluster surveys \citep[see e.g.][]{finoguenov+10}. The
evolution of $\rho_{N_{halo}}$ as a function of redshift in each mass
range is shown in Fig. \ref{g_density}. For comparison we estimate the
comoving $\rho_{N_{halo}}(z)$ in the same mass bins also according to the Planck
cosmology based on the SZ Planck number counts \citep[red lines in
  Fig. \ref{g_density}]{PC+13}. In this cosmology, the number of
clusters and groups is higher by 0.15, 0.18 and 0.25 dex, on average,
up to $z\sim 1.5$ for low mass, high mass groups, and clusters,
respectively.

Fig. \ref{sfr_density} shows the contribution of each halo mass range
to $\sfrd(z)$. In calculating such contributions we include an error of
0.35 dex in the $\tsfrm$ derived from the $\tsfrm-z-M_{200}$ as
explained in Section \ref{sfr}. The shaded blue region shows the
global evolution of obscured $\sfrd$ in all halo masses as derived by
\citet{magnelli+13}. 
Magnelli et al. (2013) combine the obscured $\sfrd$ with the
unobscured $\sfrd$ derived by \citet{Cucciati+12} using rest-frame UV
observations. However, we do not know the contribution of the
unobscured SFRD for the group galaxy population. Thus, for a fair
comparison we consider only the contribution of the group galaxy
population to the obscured $\sfrd$, as also done in Popesso et
al. (2014). We point out, however, that the obscured $\sfrd$ dominates
the total $\sfrd$ at any redshift. Indeed, Magnelli et al. (2013)
report that the unobscured $\sfrd$ accounts only for about
$\thicksim\,$25\%, $\thicksim\,$12\% and $\thicksim\,$17\% of the
total $\sfrd$ at $z$$\,\thicksim\,$$0$, $z$$\,\thicksim\,$$1$ and
$z$$\,\thicksim\,$$2$, respectively. In addition, there are no reasons
to assume that the same correction could be applied to the group
galaxy SFRD, since it is not known whether the evolution of the mean
rest-frame UV dust attenuation is environment dependent.
For completeness we overplot also the compilation
of values of \citet{Hopkins_Beacom06}, which are derived also from UV
data (black points in Fig. \ref{sfr_density}).

Fig. \ref{sfr_density} shows that the contribution of low mass groups
(with masses in the range $6{\times}10^{12}-6{\times}10^{13}$
$M_{\odot}$) provides a substantial contribution (50-80\%) to the
$\sfrd$ at z$\sim$1. Such contribution declines faster than the cosmic
$\sfrd$ between redshift 1 to the present epoch, reaching a value of
$<10\%$ at $z < 0.3$. This is consistent with our findings of Popesso
et al. (2014) based purely on the integration of the IR luminosity
function of group galaxies in four redshift intervals up to z$\sim$
1.6.

More massive systems such as high mass groups and clusters provide only
a marginal contribution ($< 10\%$ and $<1\%$, respectively) at any
epoch. This is due to two reasons: 1) they show in general a much
lower SF activity per halo mass than less massive systems and 2) their
number density is, especially at high redshift, orders of magnitude
lower that the one of low mass groups (see Fig. \ref{g_density}). In
particular, the number density of clusters is extremely low at $z > 1$
since these massive structures are created at more recent epochs. Thus,
their contribution declines at $z > 1$.

Since halos of masses larger than $\sim 6 \times 10^{12}$ provide a
negligible contribution to the cosmic $\sfrd$ at $ z < 0.3$, it
follows that this must be sustained by galaxies in lower mass
halos. This is consistent with the findings of \citet[ the former
  based on the SDSS fossil record, the latter entirely based on
  Herschel data]{heavens+04,gruppioni+13} that the most recent epoch
of the cosmic star formation history is dominated by galaxies of low
stellar masses ($M_{\star} < 10^{8-9}$ $M_{\odot}$), which most likely
inhabit very low mass halos.

We point out that the use of the Planck cosmology would not change
these conclusions.

\subsection{An alternative approach}
\label{sect:alter}
In order to understand the physical implications of the result shown
in fig. \ref{sfr_density}, we use an alternative method. If the
galaxies belonging to low mass groups are providing a substantial
contribution to $\sfrd$ at z$\sim$1 and beyond, it means that at this
epoch they must be a significant fraction of the whole galaxy
population. Our data-set does not allow us to check this possibility,
since we do not have at the moment a complete census of the galaxy
population in terms of their parent dark matter halo mass due to the
lack of sufficient spectroscopic information.

To overcome this problem, we use the predictions of the most recent
simulations to associate galaxies of a given stellar mass to their
parent halo and the observational results to match stellar mass and
SF activity. We favor, in particular, the most recent
\citet{guo+13} model since it makes use of the more recent WMAP7
cosmology rather than the WMAP1 cosmology still adopted by Kitzbichler
\& White. (2007). The most significant difference between the
cosmologies preferred by WMAP7 and WMAP1 data, is a 10\% lower value
of $\sigma_8$. This implies a lower amplitude for primordial density
fluctuations, which translates into a decrease in the number of halos
with masses above $M^*$, and an increase for those below this
characteristic mass. Thus, the model of \citet{guo+13} reproduces
rather well the clustering properties of the galaxy population of the
local Universe, and the evolution of the galaxy stellar mass function
up to high redshift. For this reason we use this simulation to match
the galaxy stellar mass to the parent halo mass.

\begin{figure*}
\begin{center}
\includegraphics[width=0.42\textwidth]{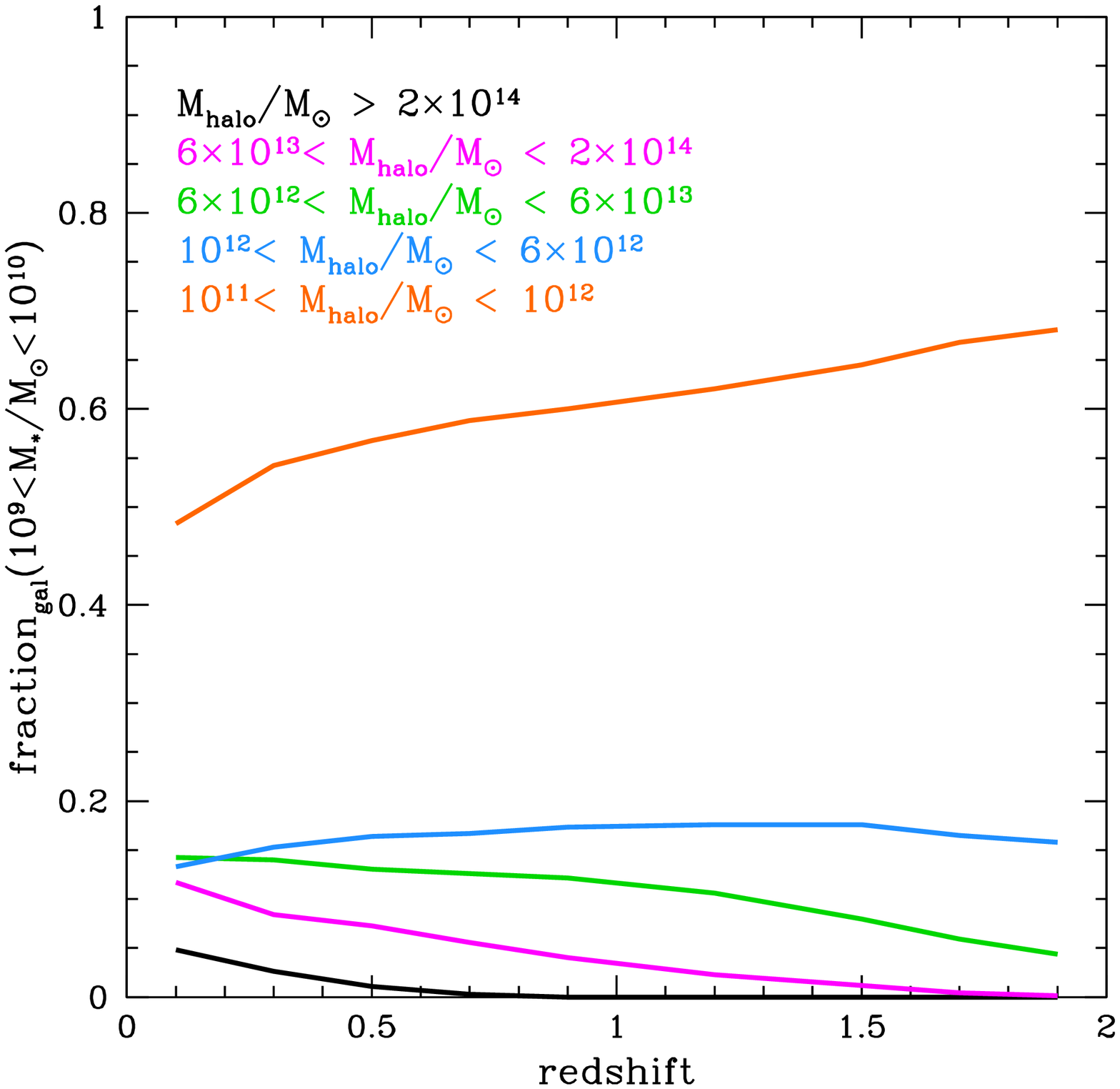}
\includegraphics[width=0.42\textwidth]{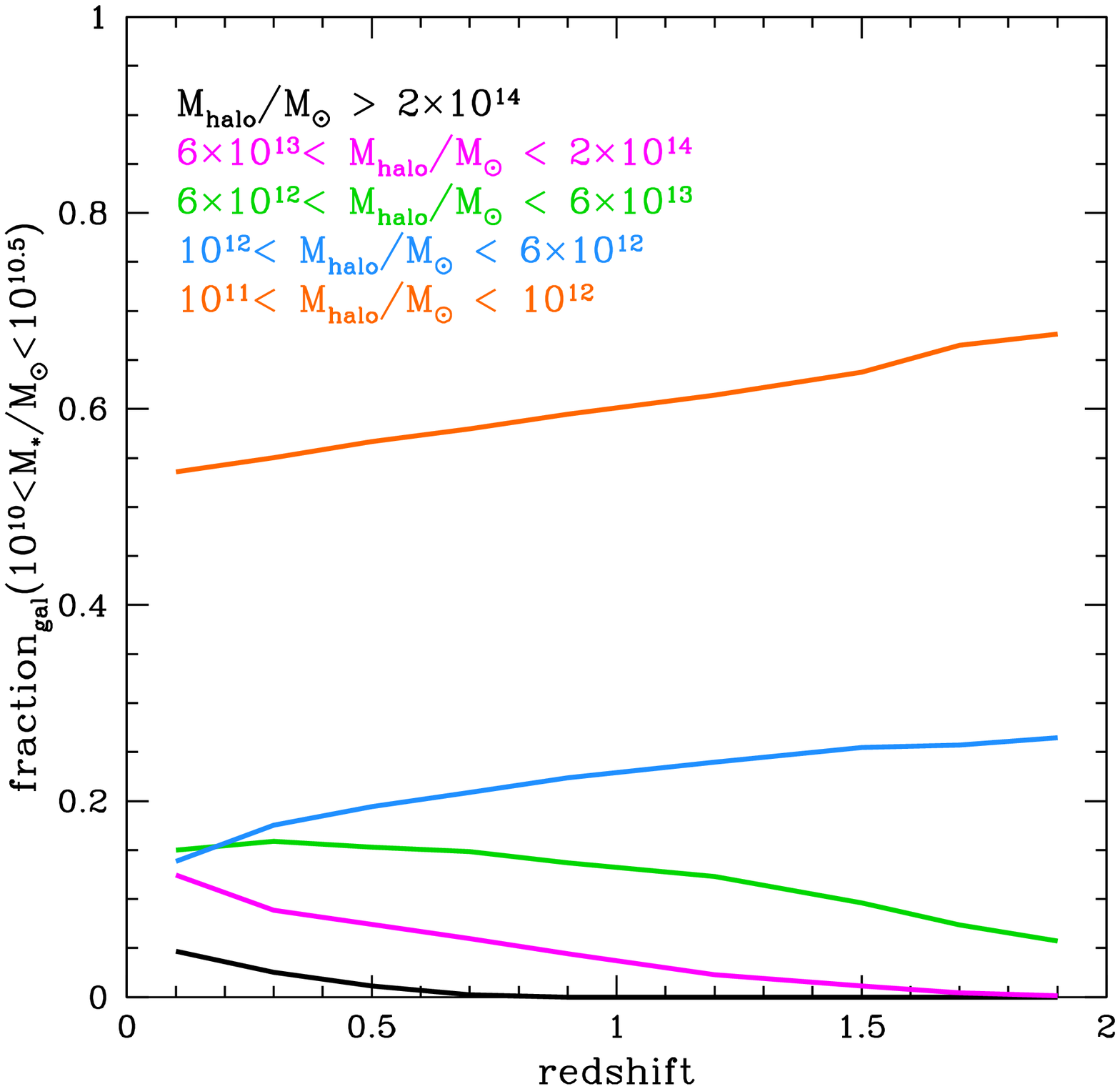}
\includegraphics[width=0.42\textwidth]{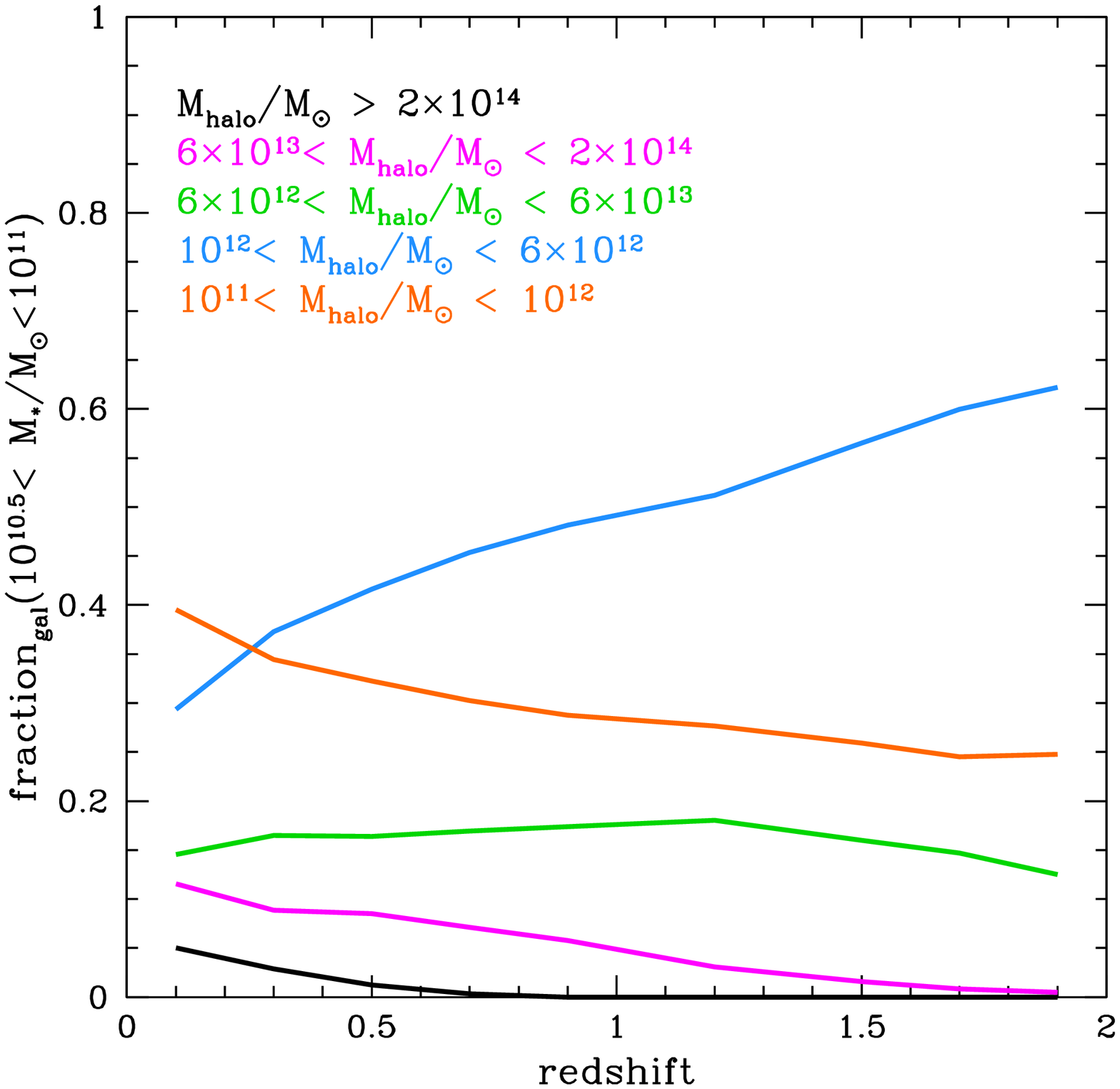}
\includegraphics[width=0.42\textwidth]{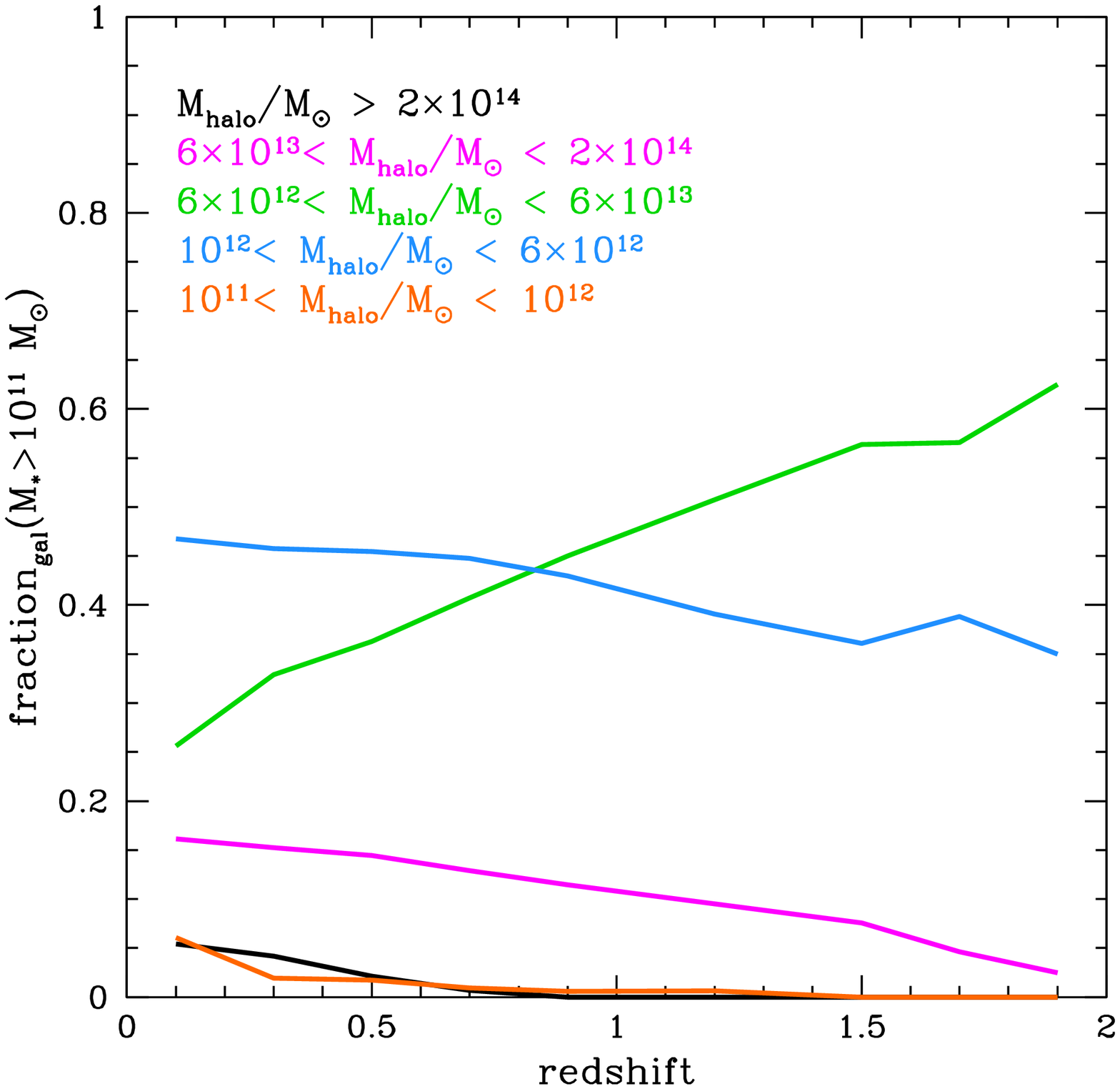} 
\caption{The fraction of galaxies belonging to parent halos of
  different masses. Each panel shows a different stellar mass range:
  $M_{\star}/M_{\odot}=10^9 - 10^{10}$ upper left, $10^{10} -
  10^{10.5}$ upper right, $10^{10.5} - 10^{11}$ bottom left, and $>
  10^{11}$ bottom right. Different colors label different parent halo
  masses, as indicated in each panel.}
\label{fraction_mhalo}
\end{center}
\end{figure*}

We use the galaxy catalogs of this model with the following
approach. We define four stellar mass ranges ( $\log (M_{\star}/
M_{\odot})=9 - 10$, $10-10.5$, $10.5 - 11$ and $> 11$) and we
estimate, in each range, the fraction of galaxies belonging to parent
halos of different mass ranges ($M_{200}/ M_{\odot}=10^{11} -
10^{12}$, $10^{12} -6{\times}10^{12}$,
$6{\times}10^{12}-6{\times}10^{13}$,
$6{\times}10^{13}-2{\times}10^{14}$,
$>2{\times}10^{14}$). Fig. \ref{fraction_mhalo} shows the redshift
evolution of these fractions and, thus, which halo mass range is
dominating in each stellar mass range. Halos with masses corresponding
to the low mass groups analyzed in this paper (masses in the range
$6{\times}10^{12}-6{\times}10^{13}$ $M_{\odot}$) dominate only at
$M_{\star}>10^{11} M_{\odot}$, and only at $z>1$. More massive halos
host only a marginal fraction $< 10\%$ of the whole galaxy population
at any stellar mass.

However, the \citet{guo+13} model shares the same problems of previous
models with respect to the level of galaxy SF activity. As discussed
in more details in next Section, also this model under predicts the
galaxy SFR as a function of $M_{\star}$ at any epoch. Thus, to link
the galaxy $M_{\star}$ to the SFR we use real data and, in particular
the observed SFR-$M_{\star}$ plane. In particular, we use the
photometric COSMOS Spitzer$+$PACS galaxy catalog matched to the
\citet{Ilbert+10} photometric redshift and $M_{\star}$ catalog. We
must limit our analysis to the LIRG regime at IR luminosities higher
than $10^{11}$ $L_{\odot}$ (SFR $> 17 M_{\odot}/yr$ according to the
Kennicutt relation 1998) and to $M_{\star} > 10^9 M_{\odot}$ to ensure
the highest photometric completeness at least up to $z\sim 1-1.2$
\citep[see][for a complete discussion about
  completeness]{Ilbert+10,magnelli+11}.

\begin{figure}
\begin{center}
\includegraphics[width=0.49\textwidth]{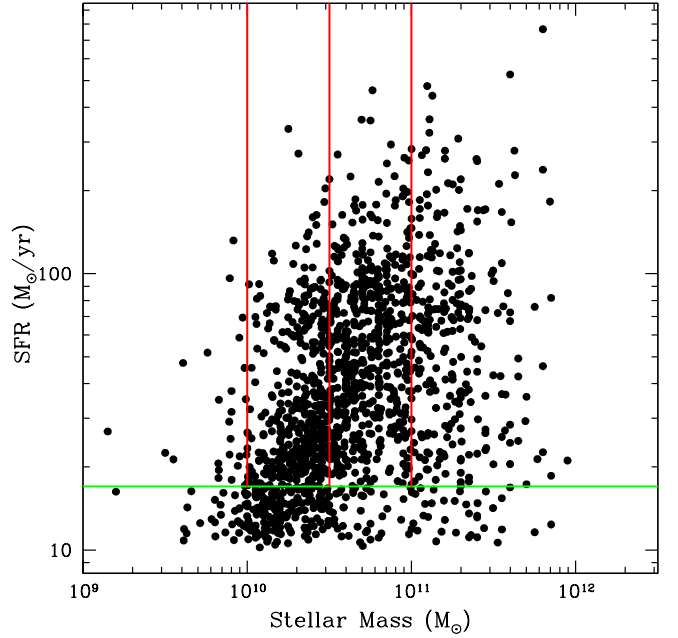} 
\caption{ Example of the SFR-stellar mass plane at $z\sim 0.5$ drawn
  from the COSMOS Spitzer$+$PACS galaxy catalog matched to the
  \citet{Ilbert+10} photometric redshift and stellar mass catalog. For
  any redshift bin, we divide the LIRG region of the SFR-stellar mass
  plane ($SFR > 17 M_{\odot}/yr$, above the green solid line) in four
  regions (red solid lines) according to the stellar mass bins defined
  in the text.}
\label{sfr_mass_plane}
\end{center}
\end{figure}

In any redshift bin, we divide the LIRG region of the SFR-$M_{\star}$
plane in four regions according to the $M_{\star}$ bins defined above
(see e.g. Fig. \ref{sfr_mass_plane}). Given the completeness of the
sample, we can calculate the fraction of total SFR due to each
$M_{\star}$ bin with respect to the total SFR of the whole LIRG
population. This is done in several redshift bins.

We combine these different estimates to calculate the fraction of the
total SFR due to each DM halo mass range limited to the LIRG
population in the following way:
\begin{equation}
{\rm SFR}(M_{halo},z)/{\rm SFR}(z)=\sum_j(f(M_{\star,j},z) \cdot f(M_{\star,j}, M_{halo}, z))
\end{equation}
where SFR$(M_{halo},z)/$SFR$(z)$ is the fraction of the total SFR due
to the LIRG population of galaxies per halo mass ($M_{halo}$) at the
redshift $z$, $f(M_{\star,j},z)$ is the fraction of the total SFR due
to the LIRGs in the $j^{th}$ stellar mass bin at redshift z and
$f(M_{\star,j}, M_{halo}, z)$ is the fraction of galaxies in the
$j^{th}$ stellar mass bin at redshift z and belonging to dark matter
halos of mass $M_{halo}$. The sum is done over the stellar mass ranges
in a given redshift bin.  We use this fractional contributions to
estimate the $\sfrd(z)$ per halo mass in the \citet{magnelli+13} $\sfrd(z)$
limited to the LIRGs. Namely, we multiply the \citet{magnelli+13} $\sfrd(z)$
by these fractional contributions.

The main limit of this approach is that it does not take into account
gradients along the MS as a function of the halo mass. In other words,
this method does not take into account that galaxies in massive halos
could favor the regions below the SF galaxy MS, in particular
at low redshift, as shown for instance in \citet{bai+09} and
\citet{ziparo+13}. In the same way this model does not take into
account that at high redshift there should be a reversal of the
SFR-density relation at the epoch when massive galaxies at the center
of groups and clusters form the bulk of their stellar population in
strong bursts of SF activity, as predicted by models
\citep[e.g.][]{delucia+06}. In other words we assume that the galaxy
SFR distribution is independent of the halo mass at any
redshift. Nevertheless, we consider that in first approximation this
method still provide a valuable way to check the robustness of our
results.

\begin{figure}
\begin{center}
\includegraphics[width=0.49\textwidth]{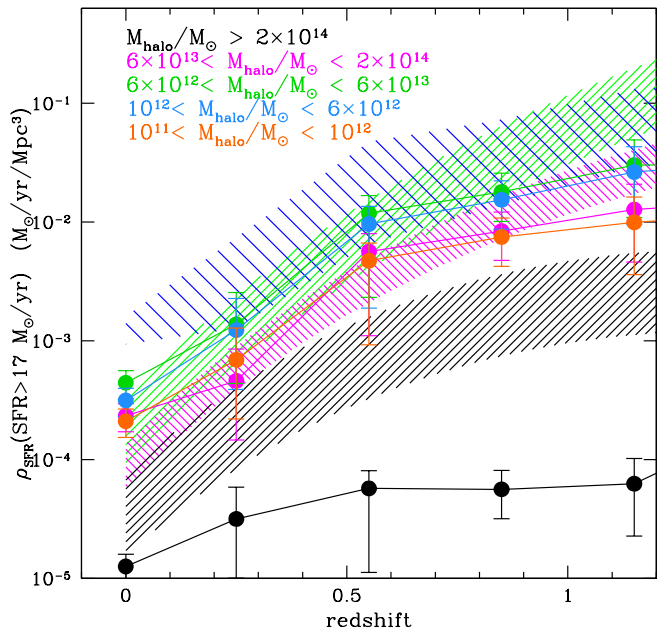} 
\caption{CSFH per halo mass limited to the LIRG population. The shaded
  regions are the CSFH per halo mass as in the left panel of
  Fig. \ref{sfr_density}. The points are the contributions per halo mass as estimated
  in the current analysis. Colors indicate different 
  halo mass ranges as indicated in the figure.}
\label{sfr_density_LIRG_check}
\end{center}
\end{figure}

The results are shown in Fig. \ref{sfr_density_LIRG_check}. The blue
shaded region is the global CSFH of Magnelli et al. (2013) limited to
the LIRG population. This is obtained by integrating the global IR
luminosity function of Magnelli et al. (2013) down to $L_{IR}=10^{11}$
$L_{\odot}$. The green, magenta and black shaded regions are the
contributions of low mass, massive groups and clusters, respectively,
to the global LIRG CSFH. These are estimated in a similar way as
described in previous section. Namely we fit the $\tsfrm-z-M_{200}$
plane by limiting the estimate of the $\tsfrm$ to the LIRG
population. The best fit relation:
\begin{equation}
\tsfrma=(0.028\pm 0.01)\times (1+z)^{9\pm0.3}\times M_{200}^{-0.47\pm0.03}
\end{equation}
is used in the same way as described in the previous section to
retrieve the contribution to the CSFH of the galaxy population
inhabiting DM halos of different mass. We point out that the evolution
of the $\tsfrm$ with redshift limited to the LIRG population is much
steeper than the relation obtained for the whole IR emitting
galaxies. This is consistent with the faster evolution of the LIRG
number and luminosity density observed already in the global
population \citep{magnelli+13,gruppioni+13} and in the same sample of
galaxy groups in Popesso et al. (2014). The halo mass dependence of
the LIRG $\tsfrm$ is, instead, quite consistent with the one of the
whole IR emitting group galaxy population. The scatter around the best
fit $\tsfrm-z-M_{200}$ plane of the LIRG population is 0.35 dex as for
the previous case.

The points in Fig. \ref{sfr_density_LIRG_check} are the contributions
per halo mass as estimated in this analysis. There is an overall good
agreement for the low mass groups (green points and shaded region) and
for the massive groups (magenta points and shaded region). There is no
agreement for the cluster mass range, but it is rather difficult to
judge if the problem is in the data due to the low number statistics
in the cluster mass regime or in the models since there are not so
many massive clusters in the Millennium Simulation due to the limited
volume. We also plot the relation obtained for the halo mass ranges
not covered by our group sample at $10^{12} < M_{halo}/M_{\odot} <
6{\times}10^{12} $ (cyan points) and $M_{halo}/M_{\odot} < 10^{12}$
(orange points).

Fig. \ref{sfr_density_LIRG_check} leads to the following
conclusions. The main contributors to the CSFH, at least for the LIRG
population, are low mass groups in the ranges $10^{12} <
M_{halo}/M_{\odot} < 6{\times}10^{12} $ and $6{\times}10^{12} <
M_{halo}/M_{\odot} < 6{\times}10^{13}$ that, together, account for
60-70\% of the CSFH at any redshift up to $z\sim 1.2$. The main reason
for this is mass segregation. Indeed, as shown in the panels of
fig. \ref{fraction_mhalo}, these groups contain the largest fraction
of massive galaxies at $M_{\star} > 10^{10.5}-10^{11}$ $M_{\odot}$, which are
highly SF along the MS. The more massive groups
and the clusters are relatively rare objects and they host only a
marginal fraction of the galaxy population at any mass, thus their
contribution to the CSFH is quite small. The DM halos in the lowest
mass range at $M_{halo}/M_{\odot} < 10^{12}$ host the majority of the
low mass galaxies, which are extremely numerous but have a  very low
SFR, according to their low mass. Thus, also these halos provide a marginal
contribution to the CSFH. If we could extend our analysis to the whole
star forming galaxy population rather than purely the LIRG, we should
likely see these halos to dominate the current epoch of the CSFH since
according to \citet{heavens+04} and Gruppioni et at (2013) low mass
galaxies are the dominant star forming galaxy population in the local
Universe.

\section{Comparison with models}
\label{comparison}

We compare here our results with several models available in the
literature in order to test their predictions.

\subsection{Semi-analytical models}
As a first approach we use the Millennium Simulation
\citep{springel+05}, which is publicly available, to perform directly
on the simulated data-sets the same analysis applied to our real
data-set. We test here the predictions of the simulated data-sets
provided by different semi-analytical models available in the
Millennium Database \citep{delucia+06,bower+06,KW07,guo+11}.

\begin{figure}
\begin{center}
\includegraphics[width=0.49\textwidth]{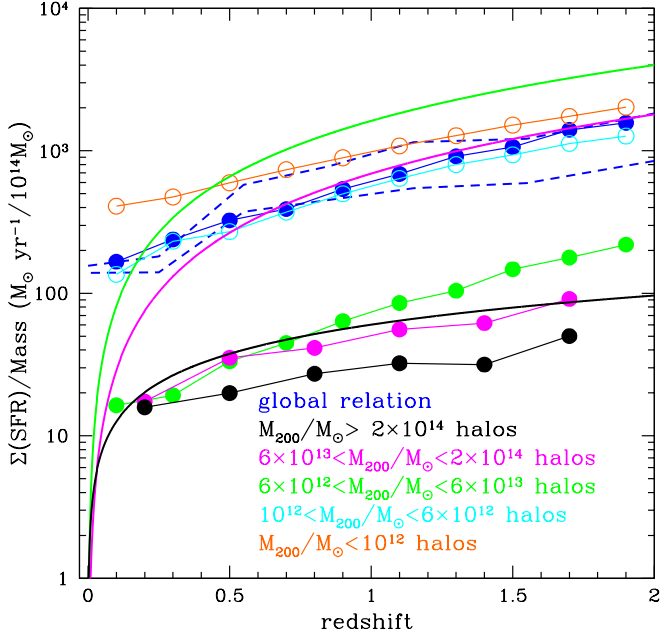} 
\caption{$\tsfrma$- redshift relation obtained from the semi-analytical
  model of \citet{delucia+06} applied to the Millennium simulation (filled and empty points).
  Different colors correspond to halos of different masses, as
  indicated in the figure. The region contained within the dashed blue line indicate the global relation obtained from the global SFR density evolution of Magnelli et al. (2013) as shown in the right panel of Fig. \ref{fig_1}. We plot here the global $\tsfrma$- redshift relation that takes into account that not all the mass is locked in halos. The solid lines show the observed best fit $\tsfrma$- redshift relations shown in the right panel of Fig. \ref{fig_1}. }
\label{tsfrm_milli}
\end{center}
\end{figure}

Fig. \ref{tsfrm_milli} shows the $\tsfrma$-redshift relation for
different halo mass ranges based in particular on the
\citet{delucia+06} model. We use $M_{200}$ as estimate of the total
mass of dark matter halos in the simulation as it is calculated in a
consistent way with respect to the observations. As in the observed
data-set, we estimate the \tsfr of each halo as the sum of all members,
identified with the same ID number by the FoF algorithm
applied by \citet{delucia+06}, and within $r_{200}$ from the central
galaxies (identified as a $type=0$ galaxy in the Millennium database).

Fig. \ref{tsfrm_milli} shows that the global $\tsfrma$-redshift
relation (blue points) is in agreement with the observations if we
consider that not all the mass is locked in halos and we correct the
global relation obtained from the CSFH of Magnelli et al. (2013) by
0.4 dex (region within the dashed lines), as derived by using the
results of \citet{Faltenbacher+10}. Nevertheless, the analysis of the
$\tsfrma$-redshift relation in different halo mass ranges, shows that
the model of \citet{delucia+06} strongly under-predicts the mean level
of activity per halo mass of all massive halos in the same range
considered in this work. According to the model the galaxy population
of halos with masses above $6\times 10^{12}$ $M_{\odot}$ lie more than
one order of magnitude below the global relation and showing a
discrepancy of more than two orders of magnitude with respect to the
observations.

\begin{figure}
\begin{center}
\includegraphics[width=0.49\textwidth]{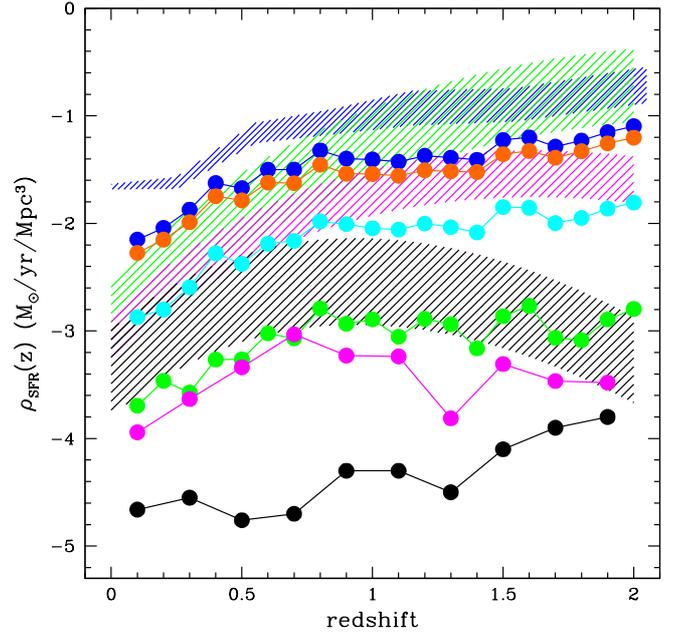} 
\caption{Contribution of halos of different mass to the CSFH as
  obtained from the semi-analytical model of \citet{delucia+06}
  applied to the Millennium simulation (filled points).  Different
  colors correspond to halos of different masses. The color coding is
  the same as in Fig. \ref{tsfrm_milli}. The shaded green, magenta,
  and black regions show the observed contribution of low mass group,
  massive group, and cluster, galaxy population, respectively to the
  global relation of Magnelli et al. (2013, blue shaded region).}
\label{sfr_density_milli}
\end{center}
\end{figure}

The models of \citet{bower+06} and \citet{guo+11} lead to very similar
results despite some marginal quantitative differences. Indeed, the
models of galaxy evolution available in the Millennium database
\citep{delucia+06,bower+06,KW07,guo+11} all predict a faster than
observed evolution of galaxies in massive halos.  This class of models
assumes that, when galaxies are accreted onto a more massive system,
the associated hot gas reservoir is stripped instantaneously. This, in
addition to the AGN feedback, induces a very rapid decline of the star
formation histories of satellite and central galaxies, respectively,
and contributes to create an excess of red and passive galaxies with
respect to the observations \citep{wang+07}. This is known as the
"over-quenching problem" for satellites galaxies. Over 95\% of the
cluster and group galaxies within the virial radius in the local
simulated Universe are passive \citep{guo+11}, at odds with
observations \citep{hansen+09,Popesso+05}. The first consequence is
that the predicted CSFH is too low with respect to observations and it
does not show the observed plateau between redshift 1 and 2 but a peak
at z$\sim$2 and a rapid decline afterwards \citep[see Fig.9
  in][]{KW07}. The second consequence is that the contribution of
group and cluster galaxies to the CSFH is
always negligible since the SF is immediately quenched (see
Fig. \ref{sfr_density_milli}). The SFR density at any epoch is
dominated by the activity of galaxies in low mass halos ($M_{halo} <
10{12} M_{\odot}$. Moreover, according to \citet{delucia+12} group
galaxies are quenched at early epochs, even before they enter the
cluster environment \citep[the pre-processing scenario,
  see][]{ZM98let}. This is at odds with the much higher level of SF
activity in low and high mass groups at any redshift with respect
to the clusters, as shown in the left panel of Fig. \ref{fig_1}.

Despite the use of the WMAP7 cosmology, which shifts the peak in cosmic star
formation rate to lower redshift, also the \citet{guo+13} model
leads qualitatively to the same results of Fig. \ref{tsfrm_milli} and
\ref{sfr_density_milli} based on \citet{delucia+06} model. The results
based on the Millennium simulation are qualitatively in agreement also
with the \citet{vandevoort+11} model, based
on a completely different set of simulations, which find that dark matter
halos with masses above $10^{13}$ $M_{\odot}$ do not contribute at all
to the cosmic star formation history of the Universe.

\subsection{Abundance matching methods}

In alternative to semi-analytical models, models using the merger
trees of hydrodynamical simulations and a conventional abundance
matching method to associate galaxies to DM halos are often used to
study the mass accretion history of galaxies as a function of their
parent halo mass \citep[e.g.]{Vale+04,Conroy+09,Behroozi+10}. It is
rather instructive to compare three of such models, which are quite
similar in the concept, but rather different in their treatment of
the mass accretion of satellite galaxies after they enter a massive
halo.

\begin{figure}
\begin{center}
\includegraphics[width=0.49\textwidth]{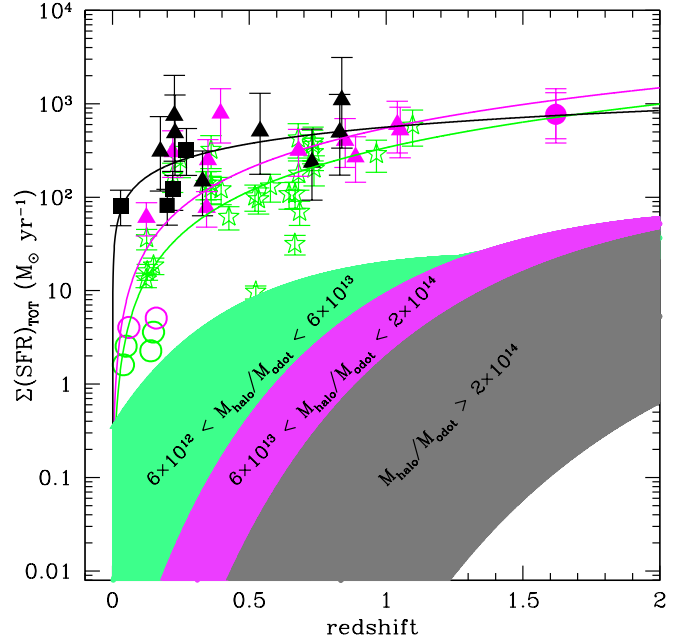} 
\caption{$\Sigma$(SFR)- redshift relation obtained from the MEAM model
  of Moster et al. (2013). The symbols and solid lines in the figure
  have the same meaning as in the left panel of Fig. \ref{fig_1}. The
  shaded regions indicates the evolution of the SFR in DM
  halos with mass at redshift $\sim 0$ in the range $6\times 10^{12} <
  M_{halo}/M_{\odot} < 6 \times 10^{13}$ (light green), $6\times
  10^{13} < M_{halo}/M_{\odot} < 2 \times 10^{14}$ (light magenta),
  and $M_{halo}/M_{\odot} > 2 \times 10^{14}$ (gray)}
\label{moster}
\end{center}
\end{figure}

The \citet{Moster+13} multi-epoch abundance matching (MEAM) model
employs a redshift-dependent parametrization of the stellar-to-halo
mass relation to populate halos and subhalos in the Millennium
simulations with galaxies, requiring that the observed stellar mass
functions at several redshifts be reproduced
simultaneously. Interestingly the model assumes that the stellar mass
of a satellite does not change after its subhalo entered the main
halo, i.e. it neglects stellar stripping and star formation in the
satellites. Thus, by construction this model implements the same
``satellite over-quenching'' observed in the Millennium simulation. In
other words, the star formation activity of any halo is located only
in its central galaxy. Moster et al. (2013) provide useful fitting
functions to estimate the redshift evolution of the SFR of a DM halo
of a given mass at redshift $\sim 0$. Fig. \ref{moster} shows the
comparison between our estimate of the $\Sigma$(SFR)- redshift
relation with the evolution of the $\Sigma$(SFR) in halos of similar
mass at redshift $\sim 0$. The comparison is not completely
straightforward as also the halo mass evolves with redshift according
to the DM halo accretion history. However, according to Moster et
al. (2013) a typical massive halo of $M_{halo}=10^{14} M_{\odot}$ at
$z \sim 0$ has grown from a $z \sim 1$ halo with a virial mass of
$M_{halo}=10^{13.6} M_{\odot}$, while lower mass halos accrete even
less in the same amount of time. Thus, the considered massive systems
at $z \sim 0$ remain in the same halo mass bin for the most of the
time window considered here. As expected, the ``satellite
over-quenching'' implemented in the model provide results consistent
with the semi-analytical model, under-predicting the relative
contribution of DM halos of different masses to the CSFH. Interestingly
enough, the model is anyhow able to reproduce the evolution of the
galaxy stellar mass function and the global CSFH.

Differently from Moster et al. (2013), \citet{Yang+12} assume that a
galaxy after becoming a satellite can gain stellar mass due to star
formation and suffer mass loss due to passive evolution. The satellite
evolution is modeled as
\begin{equation}
m_s(z)=(1-c) m_{*,a}+c m_{*,z}
\end{equation}
where $m_s(z)$ is the satellite stellar mass at redshift $z$,
$m_{*,a}$ is the mass of the satellite at the accretion time and
$m_{*,z}$ is the expected median stellar mass of central galaxies in
halos of the same mass of the satellite subhalo at redshift $z$. For
$c=0$ the satellite does not increase the mass after accretion into
the host halo as in Moster et al. (2013). Instead, for $c=1$ the
satellite accretes stellar mass in the same way as a central galaxy of
equal mass. The $c$ parameter is left free in the fitting. In addition
to the consistency with the evolution of the galaxy stellar mass
function, the best fit model is required to reproduce also the local
conditional galaxy stellar mass function (as a function of the halo
mass) of Yang et al. (2007) and the 2-point correlation function of
SDSS galaxies.  The best fit value is $c=0.98$ implying that
satellites accrete considerable mass after accretion, similarly
to the central galaxies of similar halos. According to the stellar
mass assembly histories of Yang et al. (2012), the mass increase of
such galaxies is dominated by in situ SF rather than
accretion, until the halo reaches a mass of $\sim 10^{12}$
$M_{\odot}$. This feature seems to hold independently of the final host
halo mass of the central galaxy. Thus, satellite in subhalos with
masses below $\sim 10^{12}$ $M_{\odot}$ should considerably contribute
to the overall SF activity of massive
halos. Unfortunately Yang et al. (2012) do not provide predictions for
the SF history of galaxies as a function of the host halo. Thus, a
quantitative comparison with our results is not possible, though the
qualitative predictions could lead to a better consistency with our
observations.

A step forward is done in the model of \citet{bethermin+13}.
\citet{bethermin+13} use the stellar mass function of star forming
galaxies and passive galaxies of \citet{Ilbert+10} and the halo mass
function of \citet{Tinker+08} to populate the $n^{th}$ most massive
halo with the $n^{th}$ most massive galaxy by taking into account an
increasing fraction of passive galaxies as a function of the galaxy
stellar mass. Since more massive galaxies inhabit more massive halos,
this leads naturally to a higher fraction of passive galaxies in
massive halos. The link between stellar mass and SF activity as a
function of time is done by considering the evolution of the MS of SF
galaxies. The Elbaz et al. (2011) SED templates for MS and SB galaxies
are used to estimate the mean infrared emissivity of the galaxy
population of a given halo. The model for the emissivity takes into
account a satellite quenching that is modeled as a function of the
satellite stellar mass (mass quenching) or as a function of the host
halo mass (environment quenching). The free parameters are constrained
by requiring the fit of the power spectra of the cosmic infrared
background (CIB), the cross-correlation between CIB and cosmic
microwave background lensing, and the correlation functions of bright,
resolved infrared galaxies. Though the model with the environment
quenching is better in agreement with the observational constraints,
also the mass quenching provide a reasonable fit.
\citet{bethermin+13} use the best fit model to predict the
contribution of halos in different mass ranges to the
CSFH. 

\begin{figure}
\begin{center}
\includegraphics[width=0.49\textwidth]{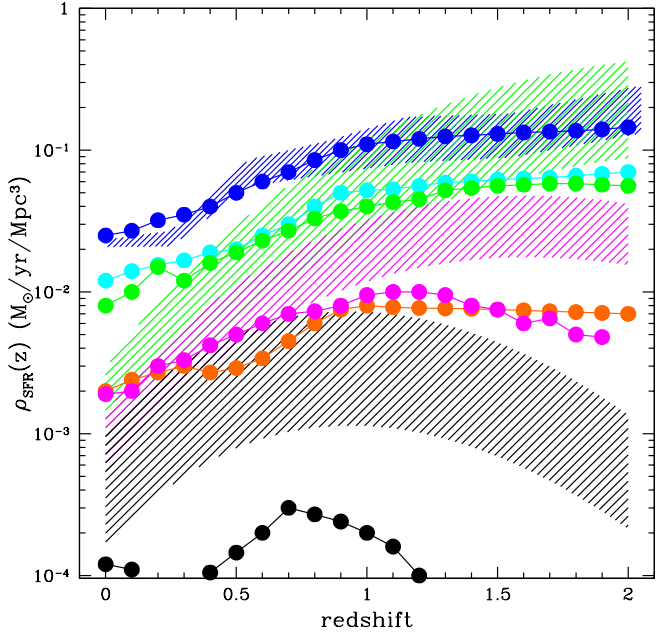} 
\caption{Contribution of halos of different mass to the CSFH as
  obtained from the abundance matching model of Bethermin et
  al. (2013, filled points).  The color coding is the same adopted in
  Fig. \ref{sfr_density_milli}. The shaded green, magenta, and black
  regions show the observed contribution of low mass group, massive
  group, and cluster, galaxy population, respectively, to the global
  relation of Magnelli et al. (2013, blue shaded region).}
\label{sfr_density_bet}
\end{center}
\end{figure}

Fig. \ref{sfr_density_bet} shows the comparison between the
\citet{bethermin+13} best model and our results. With respect to the
semi-analytical models and the Moster et al. (2013) models, the lack
of the immediate suppression of the satellite SF activity after the
accretion into the host halo, moves the bulk of the star formation
from very low mass halos ($M_{halo} < 10^{12}$ $M_{\odot}$) to more
massive halos ($10^{12} < M_{halo}/M_{\odot} < 10^{12.5}$ cyan points,
and $10^{12.5} < M_{halo}/M_{\odot} < 10^{13.5}$ green points) much
more in agreement with our results and, in particular, with the
results based on our alternative method (see
Sect.~\ref{sect:alter}). The green curve, in particular, which shows
the contribution of halos in a mass range quite consistent with our
low mass groups is consistent with our result between $0.3 < z <
1.3$. At lower redshift the SF activity is predicted to be much higher
than the observations, while at higher redshift the SFR density of
galaxies in such halos is under-predicted with respect to the green
shaded region, which at this redshift is however just an extrapolation
from our best fit $\Sigma$(SFR)$- M_{200}$-redshift relation. The
prediction of the contribution of halos in a mass range consistent
with out massive groups is consistent with the observations only up to
$z \sim 0.8$. Beyond this redshift the magenta curve is much below the
shaded region of the same color. For clusters, instead, the prediction
is largely under-estimated. We point out that the discrepancy could
arise from constraining the fraction of quenched galaxies with the
galaxy stellar mass functions of quenched and active galaxies of Ilbert
et al. (2010), which are based on SED types chosen among a few
templates only. The SED fitting technique provides a very poor constraint of
the galaxy SF activity. Indeed, as shown by \citet{Ziparo+14}, the
SFR predicted by the SED technique correlates with the more accurate
SFR derived from IR data with large scatter (0.6-0.7 dex).

\subsection{Hydrodynamics simulations}

A common feature of all previous models is that the relation between
the central galaxy stellar mass and the halo mass reaches a maximum at
halo masses $\sim 10^{12}$ $M_{\odot}$. According to Yang et
al. (2012), below this threshold the mass accretion of the central
galaxy is dominated by star formation. Thus, when the halo mass
reaches $\sim 10^{12}$ $M_{\odot}$ a process takes place to quench the
star formation. Interestingly, this mass scale is very similar to the
cold-mode to hot-mode transition scale \citep{Birnboim+03,keres+05} in
the theory of gas accretion, as derived in hydrodynamics simulations,
whereas large halos primarily accrete hot gas and low mass halos
primarily accrete cold gas. This would suggest that the quenching of
central galaxies coincides with the formation of a hot gaseous halo,
and, thus, with a lack of cold gas supply. What would be the fate of
satellites? According to \citet{simha+09}, also the subhalos retain
their identity for quite some time after accreting a larger halo. So
satellites in subhalos less massive than $\sim 10^{12}$ $M_{\odot}$ do
not immediately see the effect of the hot gas in the larger halo and
accrete in cold mode. Thus, consistently with the results of Yang et
al. (2012) and \citet{bethermin+13}, satellite galaxies continue to
accrete gas and convert it to stars over a rather large period, that
according to \citet{simha+09} is about of 0.5-1 Gyr after the
merger. The gas accretion declines steadily over this period. Since
star formation follows mass accretion with a short delay, satellites
should experience quenching in a similar amount of time. This scenario
would be consistent with our observations. Indeed, at $z\sim 1$ when
massive halos are just forming via merger, the SF activity in the
accreted subhalos is still high. At later epochs, instead, the
transition to the hot mode accretion of the satellites and the
consequent progressive quenching of their SF activity, would lead to
the faster decline of their contribution to the CSFH with respect to
lower mass halos, which evolve in a cold mode accretion phase
maintaining a high SFR.

\begin{figure}
\begin{center}
\includegraphics[width=0.49\textwidth]{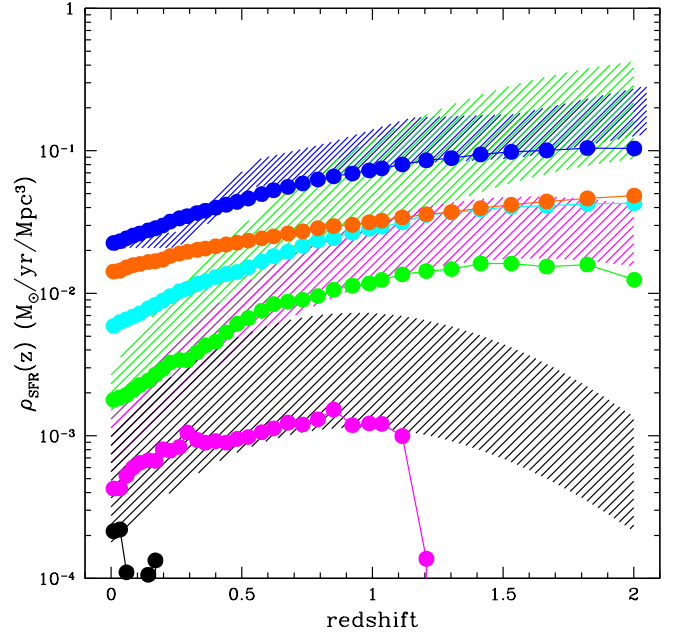} 
\caption{Contribution of halos of different mass to the CSFH as
  obtained from the {\it{Illustris}} hydrodynamical simulation (Genel et al. 2014, filled points).  The color coding of shaded regions and points is the same adopted in
  Fig. \ref{sfr_density_milli} and Fig. \ref{sfr_density_bet}.}
\label{illustris}
\end{center}
\end{figure}

Fig. \ref{illustris} shows the comparison between our observations and the predictions of the {\it{Illustris}} hydrodynamical simulation \citet{genel+14}. The simulation still underpredicts the contribution of massive halos to the CSFH. However, the underestimation of such contribution is reduced with respect to the results of semi-analytical models. This holds, in particular, for halos in the low mass group range (geen shaded region and points in the Fig. \ref{illustris}). Indeed, for this class of halos the observed and the predicted contributions to the CSFH are in agreement up to $z\sim 0.3$ and differ by $\sim$0.4 to 0.8 dex up to $z\sim 1.5$. Thus, a more physical treatment of the satellite quenching leads to an inprovements of the predictions if compared to the almost 2 orders of magnitudes disagreement of the predictions of semi-analytical models (Fig. \ref{sfr_density_milli}). However, we still see a large disagreement for more massive halos such as high mass groups (magenta shaded region and points) and clusters (black shaded region and points). We speculate that one possible explanation for such disagreement could be due to an over-suppression of the star formation activity in the massive galaxies hosted by such halos due to ``radio-mode'' AGN feedback. This hypothesis is supported by the evidence that in the  {\it{Illustris}} simulation the redshift $\sim$ 0  halos of $\sim 10^{13}$ $M_{\odot}$ are almost devoid of gas as a result of radio-mode AGN feedback, in disagreement with observations \citet{genel+14}.

\citet{cen11} proposed that not only the halo mass, but also the halo
environment on intermediate scales ($\sim$2 Mpc) could be a
determinant of the gas accretion mode. The overall heating of cosmic
gas due to the formation of large halos and large-scale structures
causes a progressively larger fraction of halos to inhabit regions
where the gas has too high an entropy to cool and continue feeding the
galaxies. Thus, the lack of cold gas supply would affect not
only halos above a give mass threshold but all halos inhabiting
overdense regions. The effect is differential in that overdense
regions are heated earlier and to higher temperatures than lower
density regions at any given time. Because larger halos tend to reside
in more overdense regions than smaller halos, the net differential
effects would naturally lead to both the standard galaxy downsizing
effect and the halo downsizing effect shown in Fig. \ref{fig_1}.

\section{Discussion and Conclusions}
\label{conclusion}
Our results can be summarized as follows:

\begin{itemize}

\item [-] the analysis of the $\tsfrma$--redshift relation in DM halos
  of different masses shows that low mass groups (halos with masses in
  the range $6{\times}10^{12}-6{\times}10^{13}$ $M_{\odot}$) lie well
  above the global (mean) relation, showing that a lot of star
  formation activity is confined in a very small volume in the most
  common and numerous group-sized dark matter halos. More massive
  groups tend to have a SF activity per halo mass consistent with the
  global relation, even if we account for an under-estimation of the
  global $\tsfrm$ of 0.4 dex \citep{Faltenbacher+10}.  Star formation
  activity is largely suppressed in the most massive, cluster-size,
  halos at any redshift.

\item [-] The analysis of the contribution to the CSFH by halos in
  different mass ranges shows that low mass groups provide a 60-80\%
  contribution to the $\sfrd$ at z$\sim$1. Such contribution declines
  faster than the cosmic $\sfrd$ between redshift 1 to the present
  epoch. This is consistent with our findings based purely on the
  integration of the IR LF of group galaxies in four redshift
  intervals up to z$\sim$ 1.6 (Popesso et al. 2014). More massive
  systems such as massive groups and clusters provide only a marginal
  contribution ($< 10\%$ and $<1\%$, respectively) at any
  epoch. Since halos of masses larger than $\sim 6 \times 10^{12}$
  $M_{\odot}$ provide a negligible contribution to the cosmic $\sfrd$
  at $ z < 0.3$, it follows that this must be entirely contributed by
  galaxies in lower mass halos, which statistically contain low mass
  galaxies \citep[this result is consistent with the findings
    of][]{heavens+04,gruppioni+13}.

\item[-] To understand our results, we use the \citet{guo+13} galaxy
  mock catalog drawn from the Millennium Simulation for matching the
  galaxy stellar mass to the galaxy host halo mass, and the observed
  SFR-stellar mass plane to associate galaxy stellar mass to SFR. This
  simple matching method allows us to check that low mass groups host
  statistically the largest fraction of massive galaxies at $z \sim
  1$. At such redshift these massive galaxies are probably still
  forming the bulk of their stellar population \citep{Rettura+10},
  making the low mass groups a significant contributor to the
  $\sfrd$. Halos at masses below $10^{12}$ $M_{\odot}$ mostly host low
  mass galaxies at any epoch, and while these are more numerous than
  the massive galaxies, they are characterized by a lower SFR. The
  most massive halos ($M_H > 10^{13.5-14} M_{\odot}$) are too rare to
  provide a non negligible contribution to the CSFH.

\item [-] The comparison of our results with the predictions of
  available models of galaxy formation and evolution shows that models
  implementing a very rapid quenching of the SF activity in satellite
  galaxies after the accretion onto massive halos
  \citep[e.g.]{delucia+06,bower+06,guo+11,guo+13,Moster+13}, fail in
  reproducing the observed level of SF activity as a function of
  redshift and halo mass. Models implementing a slower decline (on
  $\sim 1$ Gyr time scale) of the satellite SF activity (Yang et
  al. 2012, Bethermin et al. 2013) are in better agreement with our
  results.

\end{itemize}

In a previous paper \citep{Ziparo+14} we have shown, on the very same
data-set used in this paper, that the galaxy members of the low mass
groups at $z \sim 1$ are mainly MS SF galaxies, and the fraction of
quiescent galaxies, at the same redshift, is quite similar in groups
and low density regions. At lower redshift, instead, galaxies
inhabiting $\sim 10^{13}$ $M_{\odot}$ halos tend to lie below the MS
in its lower envelope or in the quiescence region, leading to a much
lower fraction of star forming galaxies in the high mass halos than in
the low mass ones.  This is confirmed by our analysis of the IR LF of
groups, also performed on the very same data-set (Popesso et
al. 2014). Indeed, groups at z$\sim 1$ host 70\% of the LIRGs and the
totality of the ULIRG population. Nearby groups, instead, contribute
less than 10\% of the global IR emitting galaxy population and do not
host the rarest and most star forming systems. This confirms that the
faster decline of the CSFH of the galaxy population inhabiting high
mass halos is due to a faster quenching of its SF activity with
respect to galaxies in lower mass halos. This quenching process must
be slow, as most of the models implementing a rapid
quenching of the SF activity in accreting satellites significantly
under-predicts the observed overall SF level of the galaxy population
in massive halos at any redshift. This is in agreement with the recent
findings of \citet{wetzel+13}. They use a group/cluster catalogs from
SDSS DR7 to study the star formation histories and quenching
timescales of satellite galaxies at $z=0$. They constrain satellite
star formation histories, finding a 'delayed-then-rapid' quenching
scenario: satellite SFRs evolve unaffected for $2-4$ Gyr after infall,
after which star formation quenches rapidly, with an e-folding time of
$< 0.8$ Gyr. This would rule out ram pressure stripping as a quenching
mechanism since this is acting on a timescale of few hundreds of Myrs
rather than Gyrs. Instead, starvation which is a slow acting process
or the delayed lack of cold gas supply implied by the cold-hot
accretion mode transition in galaxies accreting massive halos, or
maybe a combination of the two, would provide a quenching timescale
more consistent with the observations.

The evidence for a ``halo downsizing'' effect, whereby massive halos
evolve more rapidly than low mass halos \citep{NvdBD06}, fits into
this picture and it is not at odds with the current hierarchical
paradigm of structure formation. Instead, it implies that the
quenching process is driven by the accretion of galaxies from the
cosmic web into more massive halos or that the merger event that leads
to the formation of a bigger halo, causes or is followed by a further
quenching of the SF activity in the galaxy population of the building
blocks (low mass halos).

Our results point to a prominent role of the ``environmental''
quenching and, in particular of the ``satellite'' quenching in driving
the decline of the SF activity of the Universe in the last 8 billion
years. Indeed, this period coincides with the increase by more than an
order of magnitude in the number density of massive group-sized halos
and to the formation of the cosmic web as we know it. During this
structure formation process, more and more galaxies experience the
transition from central to satellite by accreting onto more massive
halos, making the low mass groups ($10^{12.5-13.5} M_{\odot}$) the
most common environment in the local Universe \citep{eke+05}. Thus, if
the massive halos are a ``SF quenching environment'', that is they
host processes able to progressively stop the SF activity,
the structure formation process itself is one of the best candidates
for driving the SF activity evolution of the Universe. This does
not exclude that AGN feedback can still play a role as an effective
quenching mechanism.  However, as shown recently by \citet{Genzel+14},
the incidence of powerful outflows in $z\sim 2$ star forming galaxies
drops dramatically below stellar masses of $10^{11} M_{\odot}$,
and the incidence of AGN in such massive galaxies is $\sim 50\%$. At
such redshift such massive galaxies represent the very high mass end
of the galaxy stellar mass function and they are likely to become the
red and dead behemoth of the local Universe. Thus, if AGN or stellar
feedback can act as a quenching process for the most massive galaxies,
another process, efficient at any stellar mass scale must be advocated
to explain the decline of the SF activity of the bulk of the galaxy
population since $z \sim 1$. Environmental quenching, as discussed
in this work, satisfies this requirement.

\begin{acknowledgements}
The authors thank G. Zamorani for the very useful comments on an early
draft of this paper.  PACS has been developed by a consortium of
institutes led by MPE (Germany) and including UVIE (Austria); KUL,
CSL, IMEC (Belgium); CEA, OAMP (France); MPIA (Germany); IFSI,
OAP/AOT, OAA/CAISMI, LENS, SISSA (Italy); IAC (Spain). This
development has been supported by the funding agencies BMVIT
(Austria), ESA-PRODEX (Belgium), CEA/CNES (France), DLR (Germany), ASI
(Italy), and CICYT/MCYT (Spain).

We gratefully acknowledge the contributions of the entire COSMOS
collaboration consisting of more than 100 scientists. More information
about the COSMOS survey is available at
http://www.astro.caltech.edu/$\sim$cosmos.

This research has made use of NASA's Astrophysics Data System, of NED,
which is operated by JPL/Caltech, under contract with NASA, and of
SDSS, which has been funded by the Sloan Foundation, NSF, the US
Department of Energy, NASA, the Japanese Monbukagakusho, the Max
Planck Society, and the Higher Education Funding Council of England.
The SDSS is managed by the participating institutions
(www.sdss.org/collaboration/credits.html).

\end{acknowledgements}

\bibliography{master}

\begin{thebibliography}{112}
\expandafter\ifx\csname natexlab\endcsname\relax\def\natexlab#1{#1}\fi

\bibitem[{{Bai} {et~al.}(2009){Bai}, {Rieke}, {Rieke}, {Christlein}, \&
  {Zabludoff}}]{bai+09}
{Bai}, L., {Rieke}, G.~H., {Rieke}, M.~J., {Christlein}, D., \& {Zabludoff},
  A.~I. 2009, \apj, 693, 1840

\bibitem[{{Bai} {et~al.}(2006){Bai}, {Rieke}, {Rieke}, {Hinz}, {Kelly}, \&
  {Blaylock}}]{Bai+06}
{Bai}, L., {Rieke}, G.~H., {Rieke}, M.~J., {et~al.} 2006, \apj, 639, 827

\bibitem[{{Barger} {et~al.}(2008){Barger}, {Cowie}, \& {Wang}}]{barger+08}
{Barger}, A.~J., {Cowie}, L.~L., \& {Wang}, W.-H. 2008, \apj, 689, 687

\bibitem[{{Behroozi} {et~al.}(2010){Behroozi}, {Conroy}, \&
  {Wechsler}}]{Behroozi+10}
{Behroozi}, P.~S., {Conroy}, C., \& {Wechsler}, R.~H. 2010, \apj, 717, 379

\bibitem[{{Berta} {et~al.}(2010){Berta}, {Magnelli}, {Lutz}, {Altieri},
  {Aussel}, {Andreani}, {Bauer}, {Bongiovanni}, {Cava}, {Cepa}, {Cimatti},
  {Daddi}, {Dominguez}, {Elbaz}, {Feuchtgruber}, {F{\"o}rster Schreiber},
  {Genzel}, {Gruppioni}, {Katterloher}, {Magdis}, {Maiolino}, {Nordon},
  {P{\'e}rez Garc{\'{\i}}a}, {Poglitsch}, {Popesso}, {Pozzi}, {Riguccini},
  {Rodighiero}, {Saintonge}, {Santini}, {Sanchez-Portal}, {Shao}, {Sturm},
  {Tacconi}, {Valtchanov}, {Wetzstein}, \& {Wieprecht}}]{Berta+10}
{Berta}, S., {Magnelli}, B., {Lutz}, D., {et~al.} 2010, \aap, 518, L30

\bibitem[{{B{\'e}thermin} {et~al.}(2013){B{\'e}thermin}, {Wang}, {Dor{\'e}},
  {Lagache}, {Sargent}, {Daddi}, {Cousin}, \& {Aussel}}]{bethermin+13}
{B{\'e}thermin}, M., {Wang}, L., {Dor{\'e}}, O., {et~al.} 2013, ArXiv e-prints

\bibitem[{{Birnboim} \& {Dekel}(2003)}]{Birnboim+03}
{Birnboim}, Y. \& {Dekel}, A. 2003, \mnras, 345, 349

\bibitem[{{Bongiorno} {et~al.}(2012){Bongiorno}, {Merloni}, {Brusa},
  {Magnelli}, {Salvato}, {Mignoli}, {Zamorani}, {Fiore}, {Rosario}, {Mainieri},
  {Hao}, {Comastri}, {Vignali}, {Balestra}, {Bardelli}, {Berta}, {Civano},
  {Kampczyk}, {Le Floc'h}, {Lusso}, {Lutz}, {Pozzetti}, {Pozzi}, {Riguccini},
  {Shankar}, \& {Silverman}}]{bongiorno+12}
{Bongiorno}, A., {Merloni}, A., {Brusa}, M., {et~al.} 2012, \mnras, 427, 3103

\bibitem[{{Bower} {et~al.}(2006){Bower}, {Benson}, {Malbon}, {Helly}, {Frenk},
  {Baugh}, {Cole}, \& {Lacey}}]{bower+06}
{Bower}, R.~G., {Benson}, A.~J., {Malbon}, R., {et~al.} 2006, \mnras, 370, 645

\bibitem[{{Capak} {et~al.}(2007){Capak}, {Aussel}, {Ajiki}, {McCracken},
  {Mobasher}, {Scoville}, {Shopbell}, {Taniguchi}, {Thompson}, {Tribiano},
  {Sasaki}, {Blain}, {Brusa}, {Carilli}, {Comastri}, {Carollo}, {Cassata},
  {Colbert}, {Ellis}, {Elvis}, {Giavalisco}, {Green}, {Guzzo}, {Hasinger},
  {Ilbert}, {Impey}, {Jahnke}, {Kartaltepe}, {Kneib}, {Koda}, {Koekemoer},
  {Komiyama}, {Leauthaud}, {Le Fevre}, {Lilly}, {Liu}, {Massey}, {Miyazaki},
  {Murayama}, {Nagao}, {Peacock}, {Pickles}, {Porciani}, {Renzini}, {Rhodes},
  {Rich}, {Salvato}, {Sanders}, {Scarlata}, {Schiminovich}, {Schinnerer},
  {Scodeggio}, {Sheth}, {Shioya}, {Tasca}, {Taylor}, {Yan}, \&
  {Zamorani}}]{Capak+07}
{Capak}, P., {Aussel}, H., {Ajiki}, M., {et~al.} 2007, \apjs, 172, 99

\bibitem[{{Caputi} {et~al.}(2007){Caputi}, {Lagache}, {Yan}, {Dole},
  {Bavouzet}, {Le Floc'h}, {Choi}, {Helou}, \& {Reddy}}]{Caputi+07}
{Caputi}, K.~I., {Lagache}, G., {Yan}, L., {et~al.} 2007, \apj, 660, 97

\bibitem[{{Cardamone} {et~al.}(2010){Cardamone}, {van Dokkum}, {Urry},
  {Taniguchi}, {Gawiser}, {Brammer}, {Taylor}, {Damen}, {Treister}, {Cobb},
  {Bond}, {Schawinski}, {Lira}, {Murayama}, {Saito}, \&
  {Sumikawa}}]{cardamone+10}
{Cardamone}, C.~N., {van Dokkum}, P.~G., {Urry}, C.~M., {et~al.} 2010, \apjs,
  189, 270

\bibitem[{{Cen}(2011)}]{cen11}
{Cen}, R. 2011, \apj, 741, 99

\bibitem[{{Cen}(2012)}]{cen+12}
{Cen}, R. 2012, \apj, 755, 28

\bibitem[{{Cimatti} {et~al.}(2008){Cimatti}, {Robberto}, {Baugh}, {Beckwith},
  {Content}, {Daddi}, {De Lucia}, {Garilli}, {Guzzo}, {Kauffmann}, {Lehnert},
  {Maccagni}, {Mart{\'{\i}}nez-Sansigre}, {Pasian}, {Reid}, {Rosati},
  {Salvaterra}, {Stiavelli}, {Wang}, {Osorio}, {Balcells}, {Bersanelli},
  {Bertoldi}, {Blaizot}, {Bottini}, {Bower}, {Bulgarelli}, {Burgasser},
  {Burigana}, {Butler}, {Casertano}, {Ciardi}, {Cirasuolo}, {Clampin}, {Cole},
  {Comastri}, {Cristiani}, {Cuby}, {Cuttaia}, {de Rosa}, {Sanchez}, {di Capua},
  {Dunlop}, {Fan}, {Ferrara}, {Finelli}, {Franceschini}, {Franx}, {Franzetti},
  {Frenk}, {Gardner}, {Gianotti}, {Grange}, {Gruppioni}, {Gruppuso}, {Hammer},
  {Hillenbrand}, {Jacobsen}, {Jarvis}, {Kennicutt}, {Kimble}, {Kriek}, {Kurk},
  {Kneib}, {Le Fevre}, {Macchetto}, {MacKenty}, {Madau}, {Magliocchetti},
  {Maino}, {Mandolesi}, {Masetti}, {McLure}, {Mennella}, {Meyer}, {Mignoli},
  {Mobasher}, {Molinari}, {Morgante}, {Morris}, {Nicastro}, {Oliva},
  {Padovani}, {Palazzi}, {Paresce}, {Garrido}, {Pian}, {Popa}, {Postman},
  {Pozzetti}, {Rayner}, {Rebolo}, {Renzini}, {R{\"o}ttgering}, {Schinnerer},
  {Scodeggio}, {Saisse}, {Shanks}, {Shapley}, {Sharples}, {Shea}, {Silk},
  {Smail}, {Span{\'o}}, {Steinacker}, {Stringhetti}, {Szalay}, {Tresse},
  {Trifoglio}, {Urry}, {Valenziano}, {Villa}, {Perez}, {Walter}, {Ward},
  {White}, {White}, {Wright}, {Wyse}, {Zamorani}, {Zacchei}, {Zeilinger}, \&
  {Zerbi}}]{Cimatti+08}
{Cimatti}, A., {Robberto}, M., {Baugh}, C., {et~al.} 2008, Experimental
  Astronomy, 37

\bibitem[{{Conroy} \& {Wechsler}(2009)}]{Conroy+09}
{Conroy}, C. \& {Wechsler}, R.~H. 2009, \apj, 696, 620

\bibitem[{{Cooper} {et~al.}(2012){Cooper}, {Yan}, {Dickinson}, {Juneau},
  {Lotz}, {Newman}, {Papovich}, {Salim}, {Walth}, {Weiner}, \&
  {Willmer}}]{cooper+12}
{Cooper}, M.~C., {Yan}, R., {Dickinson}, M., {et~al.} 2012, \mnras, 425, 2116

\bibitem[{{Cowie} {et~al.}(2004){Cowie}, {Barger}, {Fomalont}, \&
  {Capak}}]{cowie+04}
{Cowie}, L.~L., {Barger}, A.~J., {Fomalont}, E.~B., \& {Capak}, P. 2004, \apjl,
  603, L69

\bibitem[{{Cucciati} {et~al.}(2012){Cucciati}, {Tresse}, {Ilbert}, {Le
  F{\`e}vre}, {Garilli}, {Le Brun}, {Cassata}, {Franzetti}, {Maccagni},
  {Scodeggio}, {Zucca}, {Zamorani}, {Bardelli}, {Bolzonella}, {Bielby},
  {McCracken}, {Zanichelli}, \& {Vergani}}]{Cucciati+12}
{Cucciati}, O., {Tresse}, L., {Ilbert}, O., {et~al.} 2012, \aap, 539, A31

\bibitem[{{Daddi} {et~al.}(2007){Daddi}, {Dickinson}, {Morrison}, {Chary},
  {Cimatti}, {Elbaz}, {Frayer}, {Renzini}, {Pope}, {Alexander}, {Bauer},
  {Giavalisco}, {Huynh}, {Kurk}, \& {Mignoli}}]{Daddi+07}
{Daddi}, E., {Dickinson}, M., {Morrison}, G., {et~al.} 2007, \apj, 670, 156

\bibitem[{{De Lucia} {et~al.}(2006){De Lucia}, {Springel}, {White}, {Croton},
  \& {Kauffmann}}]{delucia+06}
{De Lucia}, G., {Springel}, V., {White}, S.~D.~M., {Croton}, D., \&
  {Kauffmann}, G. 2006, \mnras, 366, 499

\bibitem[{{De Lucia} {et~al.}(2012){De Lucia}, {Weinmann}, {Poggianti},
  {Arag{\'o}n-Salamanca}, \& {Zaritsky}}]{delucia+12}
{De Lucia}, G., {Weinmann}, S., {Poggianti}, B.~M., {Arag{\'o}n-Salamanca}, A.,
  \& {Zaritsky}, D. 2012, \mnras, 423, 1277

\bibitem[{{Dressler}(1980)}]{dressler+80}
{Dressler}, A. 1980, \apj, 236, 351

\bibitem[{{Eke} {et~al.}(2005){Eke}, {Baugh}, {Cole}, {Frenk}, {King}, \&
  {Peacock}}]{eke+05}
{Eke}, V.~R., {Baugh}, C.~M., {Cole}, S., {et~al.} 2005, \mnras, 362, 1233

\bibitem[{{Elbaz} {et~al.}(2007){Elbaz}, {Daddi}, {Le Borgne}, {Dickinson},
  {Alexander}, {Chary}, {Starck}, {Brandt}, {Kitzbichler}, {MacDonald},
  {Nonino}, {Popesso}, {Stern}, \& {Vanzella}}]{Elbaz+07}
{Elbaz}, D., {Daddi}, E., {Le Borgne}, D., {et~al.} 2007, \aap, 468, 33

\bibitem[{{Elbaz} {et~al.}(2011){Elbaz}, {Dickinson}, {Hwang},
  {D{\'{\i}}az-Santos}, {Magdis}, {Magnelli}, {Le Borgne}, {Galliano},
  {Pannella}, {Chanial}, {Armus}, {Charmandaris}, {Daddi}, {Aussel}, {Popesso},
  {Kartaltepe}, {Altieri}, {Valtchanov}, {Coia}, {Dannerbauer}, {Dasyra},
  {Leiton}, {Mazzarella}, {Alexander}, {Buat}, {Burgarella}, {Chary}, {Gilli},
  {Ivison}, {Juneau}, {Le Floc'h}, {Lutz}, {Morrison}, {Mullaney}, {Murphy},
  {Pope}, {Scott}, {Brodwin}, {Calzetti}, {Cesarsky}, {Charlot}, {Dole},
  {Eisenhardt}, {Ferguson}, {F{\"o}rster Schreiber}, {Frayer}, {Giavalisco},
  {Huynh}, {Koekemoer}, {Papovich}, {Reddy}, {Surace}, {Teplitz}, {Yun}, \&
  {Wilson}}]{Elbaz+11}
{Elbaz}, D., {Dickinson}, M., {Hwang}, H.~S., {et~al.} 2011, \aap, 533, A119

\bibitem[{{Erfanianfar} {et~al.}(2014){Erfanianfar}, {Popesso}, {Finoguenov},
  {Wuyts}, {Wilman}, {Biviano}, {Ziparo}, {Salvato}, {Nandra}, {Lutz}, {Elbaz},
  {Dickinson}, {Tanaka}, {Mirkazemi}, {Balogh}, {Altieri}, {Aussel}, {Bauer},
  {Berta}, {Bielby}, {Brandt}, {Cappelluti}, {Cimatti}, {Cooper}, {Fadda},
  {Ilbert}, {Le Floch}, {Magnelli}, {Mulchaey}, {Nordon}, {Newman},
  {Poglitsch}, \& {Pozzi}}]{erfanianfar+14}
{Erfanianfar}, G., {Popesso}, P., {Finoguenov}, A., {et~al.} 2014, \mnras, 445,
  2725

\bibitem[{{Faltenbacher} {et~al.}(2010){Faltenbacher}, {Finoguenov}, \&
  {Drory}}]{Faltenbacher+10}
{Faltenbacher}, A., {Finoguenov}, A., \& {Drory}, N. 2010, \apj, 712, 484

\bibitem[{{Finn} {et~al.}(2010){Finn}, {Desai}, {Rudnick}, {Poggianti}, {Bell},
  {Hinz}, {Jablonka}, {Milvang-Jensen}, {Moustakas}, {Rines}, \&
  {Zaritsky}}]{Finn+10}
{Finn}, R.~A., {Desai}, V., {Rudnick}, G., {et~al.} 2010, \apj, 720, 87

\bibitem[{{Finoguenov} {et~al.}(2010){Finoguenov}, {Watson}, {Tanaka},
  {Simpson}, {Cirasuolo}, {Dunlop}, {Peacock}, {Farrah}, {Akiyama}, {Ueda},
  {Smol{\v c}i{\'c}}, {Stewart}, {Rawlings}, {van Breukelen}, {Almaini},
  {Clewley}, {Bonfield}, {Jarvis}, {Barr}, {Foucaud}, {McLure}, {Sekiguchi}, \&
  {Egami}}]{finoguenov+10}
{Finoguenov}, A., {Watson}, M.~G., {Tanaka}, M., {et~al.} 2010, \mnras, 403,
  2063

\bibitem[{{Gavazzi} {et~al.}(2006){Gavazzi}, {O'Neil}, {Boselli}, \& {van
  Driel}}]{Gavazzi+06}
{Gavazzi}, G., {O'Neil}, K., {Boselli}, A., \& {van Driel}, W. 2006, \aap, 449,
  929

\bibitem[{{Genel} {et~al.}(2014){Genel}, {Vogelsberger}, {Springel}, {Sijacki},
  {Nelson}, {Snyder}, {Rodriguez-Gomez}, {Torrey}, \& {Hernquist}}]{genel+14}
{Genel}, S., {Vogelsberger}, M., {Springel}, V., {et~al.} 2014, \mnras, 445,
  175

\bibitem[{{Genzel} {et~al.}(2014){Genzel}, {F{\"o}rster Schreiber}, {Rosario},
  {Lang}, {Lutz}, {Wisnioski}, {Wuyts}, {Wuyts}, {Bandara}, {Bender}, {Berta},
  {Kurk}, {Mendel}, {Tacconi}, {Wilman}, {Beifiori}, {Brammer}, {Burkert},
  {Buschkamp}, {Chan}, {Carollo}, {Davies}, {Eisenhauer}, {Fabricius},
  {Fossati}, {Kriek}, {Kulkarni}, {Lilly}, {Mancini}, {Momcheva}, {Naab},
  {Nelson}, {Renzini}, {Saglia}, {Sharples}, {Sternberg}, {Tacchella}, \& {van
  Dokkum}}]{Genzel+14}
{Genzel}, R., {F{\"o}rster Schreiber}, N.~M., {Rosario}, D., {et~al.} 2014,
  ArXiv e-prints

\bibitem[{{G{\'o}mez} {et~al.}(2003){G{\'o}mez}, {Nichol}, {Miller}, {Balogh},
  {Goto}, {Zabludoff}, {Romer}, {Bernardi}, {Sheth}, {Hopkins}, {Castander},
  {Connolly}, {Schneider}, {Brinkmann}, {Lamb}, {SubbaRao}, \&
  {York}}]{Gomez+03}
{G{\'o}mez}, P.~L., {Nichol}, R.~C., {Miller}, C.~J., {et~al.} 2003, \apj, 584,
  210

\bibitem[{{Gruppioni} {et~al.}(2013){Gruppioni}, {Pozzi}, {Rodighiero},
  {Delvecchio}, {Berta}, {Pozzetti}, {Zamorani}, {Andreani}, {Cimatti},
  {Ilbert}, {Le Floc'h}, {Lutz}, {Magnelli}, {Marchetti}, {Monaco}, {Nordon},
  {Oliver}, {Popesso}, {Riguccini}, {Roseboom}, {Rosario}, {Sargent},
  {Vaccari}, {Altieri}, {Aussel}, {Bongiovanni}, {Cepa}, {Daddi},
  {Dom{\'{\i}}nguez-S{\'a}nchez}, {Elbaz}, {F{\"o}rster Schreiber}, {Genzel},
  {Iribarrem}, {Magliocchetti}, {Maiolino}, {Poglitsch}, {P{\'e}rez
  Garc{\'{\i}}a}, {Sanchez-Portal}, {Sturm}, {Tacconi}, {Valtchanov},
  {Amblard}, {Arumugam}, {Bethermin}, {Bock}, {Boselli}, {Buat}, {Burgarella},
  {Castro-Rodr{\'{\i}}guez}, {Cava}, {Chanial}, {Clements}, {Conley}, {Cooray},
  {Dowell}, {Dwek}, {Eales}, {Franceschini}, {Glenn}, {Griffin},
  {Hatziminaoglou}, {Ibar}, {Isaak}, {Ivison}, {Lagache}, {Levenson}, {Lu},
  {Madden}, {Maffei}, {Mainetti}, {Nguyen}, {O'Halloran}, {Page}, {Panuzzo},
  {Papageorgiou}, {Pearson}, {P{\'e}rez-Fournon}, {Pohlen}, {Rigopoulou},
  {Rowan-Robinson}, {Schulz}, {Scott}, {Seymour}, {Shupe}, {Smith}, {Stevens},
  {Symeonidis}, {Trichas}, {Tugwell}, {Vigroux}, {Wang}, {Wright}, {Xu},
  {Zemcov}, {Bardelli}, {Carollo}, {Contini}, {Le F{\'e}vre}, {Lilly},
  {Mainieri}, {Renzini}, {Scodeggio}, \& {Zucca}}]{gruppioni+13}
{Gruppioni}, C., {Pozzi}, F., {Rodighiero}, G., {et~al.} 2013, \mnras, 432, 23

\bibitem[{{Guo} {et~al.}(2014){Guo}, {Lacey}, {Norberg}, {Cole}, {Baugh},
  {Frenk}, {Cooray}, {Dye}, {Bourne}, {Dunne}, {Eales}, {Ivison}, {Maddox},
  {Alpasan}, {Baldry}, {Driver}, \& {Robotham}}]{Guo+14}
{Guo}, Q., {Lacey}, C., {Norberg}, P., {et~al.} 2014, ArXiv e-prints

\bibitem[{{Guo} {et~al.}(2013){Guo}, {White}, {Angulo}, {Henriques}, {Lemson},
  {Boylan-Kolchin}, {Thomas}, \& {Short}}]{guo+13}
{Guo}, Q., {White}, S., {Angulo}, R.~E., {et~al.} 2013, \mnras, 428, 1351

\bibitem[{{Guo} {et~al.}(2011){Guo}, {White}, {Boylan-Kolchin}, {De Lucia},
  {Kauffmann}, {Lemson}, {Li}, {Springel}, \& {Weinmann}}]{guo+11}
{Guo}, Q., {White}, S., {Boylan-Kolchin}, M., {et~al.} 2011, \mnras, 413, 101

\bibitem[{{Haines} {et~al.}(2013){Haines}, {Pereira}, {Smith}, {Egami},
  {Sanderson}, {Babul}, {Finoguenov}, {Merluzzi}, {Busarello}, {Rawle}, \&
  {Okabe}}]{Haines+13}
{Haines}, C.~P., {Pereira}, M.~J., {Smith}, G.~P., {et~al.} 2013, \apj, 775,
  126

\bibitem[{{Haines} {et~al.}(2010){Haines}, {Smith}, {Pereira}, {Egami},
  {Moran}, {Hardegree-Ullman}, {Rawle}, \& {Rex}}]{Haines+10}
{Haines}, C.~P., {Smith}, G.~P., {Pereira}, M.~J., {et~al.} 2010, \aap, 518,
  L19

\bibitem[{{Hansen} {et~al.}(2009){Hansen}, {Sheldon}, {Wechsler}, \&
  {Koester}}]{hansen+09}
{Hansen}, S.~M., {Sheldon}, E.~S., {Wechsler}, R.~H., \& {Koester}, B.~P. 2009,
  \apj, 699, 1333

\bibitem[{{Harrison} {et~al.}(2012){Harrison}, {Alexander}, {Mullaney},
  {Altieri}, {Coia}, {Charmandaris}, {Daddi}, {Dannerbauer}, {Dasyra}, {Del
  Moro}, {Dickinson}, {Hickox}, {Ivison}, {Kartaltepe}, {Le Floc'h}, {Leiton},
  {Magnelli}, {Popesso}, {Rovilos}, {Rosario}, \& {Swinbank}}]{harrison+12}
{Harrison}, C.~M., {Alexander}, D.~M., {Mullaney}, J.~R., {et~al.} 2012, \apjl,
  760, L15

\bibitem[{{Heavens} {et~al.}(2004){Heavens}, {Panter}, {Jimenez}, \&
  {Dunlop}}]{heavens+04}
{Heavens}, A., {Panter}, B., {Jimenez}, R., \& {Dunlop}, J. 2004, \nat, 428,
  625

\bibitem[{{Hopkins} \& {Beacom}(2006)}]{Hopkins_Beacom06}
{Hopkins}, A.~M. \& {Beacom}, J.~F. 2006, \apj, 651, 142

\bibitem[{{Ilbert} {et~al.}(2010){Ilbert}, {Salvato}, {Le Floc'h}, {Aussel},
  {Capak}, {McCracken}, {Mobasher}, {Kartaltepe}, {Scoville}, {Sanders},
  {Arnouts}, {Bundy}, {Cassata}, {Kneib}, {Koekemoer}, {Le F{\`e}vre}, {Lilly},
  {Surace}, {Taniguchi}, {Tasca}, {Thompson}, {Tresse}, {Zamojski}, {Zamorani},
  \& {Zucca}}]{Ilbert+10}
{Ilbert}, O., {Salvato}, M., {Le Floc'h}, E., {et~al.} 2010, \apj, 709, 644

\bibitem[{{Kennicutt}(1998{\natexlab{a}})}]{Kennicutt98}
{Kennicutt}, Jr., R.~C. 1998{\natexlab{a}}, \araa, 36, 189

\bibitem[{{Kennicutt}(1998{\natexlab{b}})}]{Kennicutt+98}
{Kennicutt}, Jr., R.~C. 1998{\natexlab{b}}, \araa, 36, 189

\bibitem[{{Kere{\v s}} {et~al.}(2005){Kere{\v s}}, {Katz}, {Weinberg}, \&
  {Dav{\'e}}}]{keres+05}
{Kere{\v s}}, D., {Katz}, N., {Weinberg}, D.~H., \& {Dav{\'e}}, R. 2005,
  \mnras, 363, 2

\bibitem[{{Kitzbichler} \& {White}(2007)}]{KW07}
{Kitzbichler}, M.~G. \& {White}, S.~D.~M. 2007, \mnras, 376, 2

\bibitem[{{Kurk} {et~al.}(2008){Kurk}, {Cimatti}, {Zamorani}, {Halliday},
  {Mignoli}, {Pozzetti}, {Daddi}, {Rosati}, {Dickinson}, {Bolzonella},
  {Cassata}, {Renzini}, {Franceschini}, {Rodighiero}, \& {Berta}}]{Kurk+08}
{Kurk}, J., {Cimatti}, A., {Zamorani}, G., {et~al.} 2008, in Astronomical
  Society of the Pacific Conference Series, Vol. 399, Panoramic Views of Galaxy
  Formation and Evolution, ed. {T.~Kodama, T.~Yamada, \& K.~Aoki}, 332

\bibitem[{{Le Floc'h} {et~al.}(2009){Le Floc'h}, {Aussel}, {Ilbert},
  {Riguccini}, {Frayer}, {Salvato}, {Arnouts}, {Surace}, {Feruglio},
  {Rodighiero}, {Capak}, {Kartaltepe}, {Heinis}, {Sheth}, {Yan}, {McCracken},
  {Thompson}, {Sanders}, {Scoville}, \& {Koekemoer}}]{LeFloch+09}
{Le Floc'h}, E., {Aussel}, H., {Ilbert}, O., {et~al.} 2009, \apj, 703, 222

\bibitem[{{Le Floc'h} {et~al.}(2005){Le Floc'h}, {Papovich}, {Dole}, {Bell},
  {Lagache}, {Rieke}, {Egami}, {P{\'e}rez-Gonz{\'a}lez}, {Alonso-Herrero},
  {Rieke}, {Blaylock}, {Engelbracht}, {Gordon}, {Hines}, {Misselt}, {Morrison},
  \& {Mould}}]{LeFloch+05}
{Le Floc'h}, E., {Papovich}, C., {Dole}, H., {et~al.} 2005, \apj, 632, 169

\bibitem[{{Leauthaud} {et~al.}(2010){Leauthaud}, {Finoguenov}, {Kneib},
  {Taylor}, {Massey}, {Rhodes}, {Ilbert}, {Bundy}, {Tinker}, {George}, {Capak},
  {Koekemoer}, {Johnston}, {Zhang}, {Cappelluti}, {Ellis}, {Elvis}, {Giodini},
  {Heymans}, {Le F{\`e}vre}, {Lilly}, {McCracken}, {Mellier},
  {R{\'e}fr{\'e}gier}, {Salvato}, {Scoville}, {Smoot}, {Tanaka}, {Van
  Waerbeke}, \& {Wolk}}]{Leauthaud+10}
{Leauthaud}, A., {Finoguenov}, A., {Kneib}, J.-P., {et~al.} 2010, \apj, 709, 97

\bibitem[{{Lilly} {et~al.}(2009){Lilly}, {Le Brun}, {Maier}, {Mainieri},
  {Mignoli}, {Scodeggio}, {Zamorani}, {Carollo}, {Contini}, {Kneib}, {Le
  F{\`e}vre}, {Renzini}, {Bardelli}, {Bolzonella}, {Bongiorno}, {Caputi},
  {Coppa}, {Cucciati}, {de la Torre}, {de Ravel}, {Franzetti}, {Garilli},
  {Iovino}, {Kampczyk}, {Kovac}, {Knobel}, {Lamareille}, {Le Borgne}, {Pello},
  {Peng}, {P{\'e}rez-Montero}, {Ricciardelli}, {Silverman}, {Tanaka}, {Tasca},
  {Tresse}, {Vergani}, {Zucca}, {Ilbert}, {Salvato}, {Oesch}, {Abbas},
  {Bottini}, {Capak}, {Cappi}, {Cassata}, {Cimatti}, {Elvis}, {Fumana},
  {Guzzo}, {Hasinger}, {Koekemoer}, {Leauthaud}, {Maccagni}, {Marinoni},
  {McCracken}, {Memeo}, {Meneux}, {Porciani}, {Pozzetti}, {Sanders},
  {Scaramella}, {Scarlata}, {Scoville}, {Shopbell}, \& {Taniguchi}}]{Lilly+09}
{Lilly}, S.~J., {Le Brun}, V., {Maier}, C., {et~al.} 2009, \apjs, 184, 218

\bibitem[{{Lilly} {et~al.}(1996){Lilly}, {Le Fevre}, {Hammer}, \&
  {Crampton}}]{Lilly+96}
{Lilly}, S.~J., {Le Fevre}, O., {Hammer}, F., \& {Crampton}, D. 1996, \apjl,
  460, L1

\bibitem[{{Lilly} {et~al.}(2007){Lilly}, {Le F{\`e}vre}, {Renzini}, {Zamorani},
  {Scodeggio}, {Contini}, {Carollo}, {Hasinger}, {Kneib}, {Iovino}, {Le Brun},
  {Maier}, {Mainieri}, {Mignoli}, {Silverman}, {Tasca}, {Bolzonella},
  {Bongiorno}, {Bottini}, {Capak}, {Caputi}, {Cimatti}, {Cucciati}, {Daddi},
  {Feldmann}, {Franzetti}, {Garilli}, {Guzzo}, {Ilbert}, {Kampczyk}, {Kovac},
  {Lamareille}, {Leauthaud}, {Borgne}, {McCracken}, {Marinoni}, {Pello},
  {Ricciardelli}, {Scarlata}, {Vergani}, {Sanders}, {Schinnerer}, {Scoville},
  {Taniguchi}, {Arnouts}, {Aussel}, {Bardelli}, {Brusa}, {Cappi}, {Ciliegi},
  {Finoguenov}, {Foucaud}, {Franceschini}, {Halliday}, {Impey}, {Knobel},
  {Koekemoer}, {Kurk}, {Maccagni}, {Maddox}, {Marano}, {Marconi}, {Meneux},
  {Mobasher}, {Moreau}, {Peacock}, {Porciani}, {Pozzetti}, {Scaramella},
  {Schiminovich}, {Shopbell}, {Smail}, {Thompson}, {Tresse}, {Vettolani},
  {Zanichelli}, \& {Zucca}}]{Lilly+07}
{Lilly}, S.~J., {Le F{\`e}vre}, O., {Renzini}, A., {et~al.} 2007, \apjs, 172,
  70

\bibitem[{{Lutz} {et~al.}(2011){Lutz}, {Poglitsch}, {Altieri}, {Andreani},
  {Aussel}, {Berta}, {Bongiovanni}, {Brisbin}, {Cava}, {Cepa}, {Cimatti},
  {Daddi}, {Dominguez-Sanchez}, {Elbaz}, {F{\"o}rster Schreiber}, {Genzel},
  {Grazian}, {Gruppioni}, {Harwit}, {Le Floc'h}, {Magdis}, {Magnelli},
  {Maiolino}, {Nordon}, {P{\'e}rez Garc{\'{\i}}a}, {Popesso}, {Pozzi},
  {Riguccini}, {Rodighiero}, {Saintonge}, {Sanchez Portal}, {Santini}, {Shao},
  {Sturm}, {Tacconi}, {Valtchanov}, {Wetzstein}, \& {Wieprecht}}]{Lutz+11}
{Lutz}, D., {Poglitsch}, A., {Altieri}, B., {et~al.} 2011, \aap, 532, A90

\bibitem[{{Madau} {et~al.}(1998){Madau}, {Pozzetti}, \& {Dickinson}}]{Madau+98}
{Madau}, P., {Pozzetti}, L., \& {Dickinson}, M. 1998, \apj, 498, 106

\bibitem[{{Magnelli} {et~al.}(2009){Magnelli}, {Elbaz}, {Chary}, {Dickinson},
  {Le Borgne}, {Frayer}, \& {Willmer}}]{magnelli+09}
{Magnelli}, B., {Elbaz}, D., {Chary}, R.~R., {et~al.} 2009, \aap, 496, 57

\bibitem[{{Magnelli} {et~al.}(2011){Magnelli}, {Elbaz}, {Chary}, {Dickinson},
  {Le Borgne}, {Frayer}, \& {Willmer}}]{magnelli+11}
{Magnelli}, B., {Elbaz}, D., {Chary}, R.~R., {et~al.} 2011, \aap, 528, A35

\bibitem[{{Magnelli} {et~al.}(2013){Magnelli}, {Popesso}, {Berta}, {Pozzi},
  {Elbaz}, {Lutz}, {Dickinson}, {Altieri}, {Andreani}, {Aussel},
  {B{\'e}thermin}, {Bongiovanni}, {Cepa}, {Charmandaris}, {Chary}, {Cimatti},
  {Daddi}, {F{\"o}rster Schreiber}, {Genzel}, {Gruppioni}, {Harwit}, {Hwang},
  {Ivison}, {Magdis}, {Maiolino}, {Murphy}, {Nordon}, {Pannella}, {P{\'e}rez
  Garc{\'{\i}}a}, {Poglitsch}, {Rosario}, {Sanchez-Portal}, {Santini}, {Scott},
  {Sturm}, {Tacconi}, \& {Valtchanov}}]{magnelli+13}
{Magnelli}, B., {Popesso}, P., {Berta}, S., {et~al.} 2013, \aap, 553, A132

\bibitem[{{Magorrian} {et~al.}(1998){Magorrian}, {Tremaine}, {Richstone},
  {Bender}, {Bower}, {Dressler}, {Faber}, {Gebhardt}, {Green}, {Grillmair},
  {Kormendy}, \& {Lauer}}]{magorrian+98}
{Magorrian}, J., {Tremaine}, S., {Richstone}, D., {et~al.} 1998, \aj, 115, 2285

\bibitem[{{Mamon} {et~al.}(2013){Mamon}, {Biviano}, \& {Bou{\'e}}}]{Mamon+13}
{Mamon}, G.~A., {Biviano}, A., \& {Bou{\'e}}, G. 2013, \mnras, 429, 3079

\bibitem[{{Moster} {et~al.}(2013){Moster}, {Naab}, \& {White}}]{Moster+13}
{Moster}, B.~P., {Naab}, T., \& {White}, S.~D.~M. 2013, \mnras, 428, 3121

\bibitem[{{Mullaney} {et~al.}(2012){Mullaney}, {Pannella}, {Daddi},
  {Alexander}, {Elbaz}, {Hickox}, {Bournaud}, {Altieri}, {Aussel}, {Coia},
  {Dannerbauer}, {Dasyra}, {Dickinson}, {Hwang}, {Kartaltepe}, {Leiton},
  {Magdis}, {Magnelli}, {Popesso}, {Valtchanov}, {Bauer}, {Brandt}, {Del Moro},
  {Hanish}, {Ivison}, {Juneau}, {Luo}, {Lutz}, {Sargent}, {Scott}, \&
  {Xue}}]{mullaney+12}
{Mullaney}, J.~R., {Pannella}, M., {Daddi}, E., {et~al.} 2012, \mnras, 419, 95

\bibitem[{{Neistein} {et~al.}(2006){Neistein}, {van den Bosch}, \&
  {Dekel}}]{NvdBD06}
{Neistein}, E., {van den Bosch}, F.~C., \& {Dekel}, A. 2006, \mnras, 372, 933

\bibitem[{{Noeske} {et~al.}(2007){Noeske}, {Weiner}, {Faber}, {Papovich},
  {Koo}, {Somerville}, {Bundy}, {Conselice}, {Newman}, {Schiminovich}, {Le
  Floc'h}, {Coil}, {Rieke}, {Lotz}, {Primack}, {Barmby}, {Cooper}, {Davis},
  {Ellis}, {Fazio}, {Guhathakurta}, {Huang}, {Kassin}, {Martin}, {Phillips},
  {Rich}, {Small}, {Willmer}, \& {Wilson}}]{Noeske+07}
{Noeske}, K.~G., {Weiner}, B.~J., {Faber}, S.~M., {et~al.} 2007, \apjl, 660,
  L43

\bibitem[{{Nordon} {et~al.}(2010){Nordon}, {Lutz}, {Shao}, {Magnelli}, {Berta},
  {Altieri}, {Andreani}, {Aussel}, {Bongiovanni}, {Cava}, {Cepa}, {Cimatti},
  {Daddi}, {Dominguez}, {Elbaz}, {F{\"o}rster Schreiber}, {Genzel}, {Grazian},
  {Magdis}, {Maiolino}, {P{\'e}rez Garc{\'{\i}}a}, {Poglitsch}, {Popesso},
  {Pozzi}, {Riguccini}, {Rodighiero}, {Saintonge}, {Sanchez-Portal}, {Santini},
  {Sturm}, {Tacconi}, {Valtchanov}, {Wetzstein}, \& {Wieprecht}}]{nordon+10}
{Nordon}, R., {Lutz}, D., {Shao}, L., {et~al.} 2010, \aap, 518, L24

\bibitem[{{Peng} {et~al.}(2010){Peng}, {Lilly}, {Kova{\v c}}, {Bolzonella},
  {Pozzetti}, {Renzini}, {Zamorani}, {Ilbert}, {Knobel}, {Iovino}, {Maier},
  {Cucciati}, {Tasca}, {Carollo}, {Silverman}, {Kampczyk}, {de Ravel},
  {Sanders}, {Scoville}, {Contini}, {Mainieri}, {Scodeggio}, {Kneib}, {Le
  F{\`e}vre}, {Bardelli}, {Bongiorno}, {Caputi}, {Coppa}, {de la Torre},
  {Franzetti}, {Garilli}, {Lamareille}, {Le Borgne}, {Le Brun}, {Mignoli},
  {Perez Montero}, {Pello}, {Ricciardelli}, {Tanaka}, {Tresse}, {Vergani},
  {Welikala}, {Zucca}, {Oesch}, {Abbas}, {Barnes}, {Bordoloi}, {Bottini},
  {Cappi}, {Cassata}, {Cimatti}, {Fumana}, {Hasinger}, {Koekemoer},
  {Leauthaud}, {Maccagni}, {Marinoni}, {McCracken}, {Memeo}, {Meneux}, {Nair},
  {Porciani}, {Presotto}, \& {Scaramella}}]{Peng+10}
{Peng}, Y., {Lilly}, S.~J., {Kova{\v c}}, K., {et~al.} 2010, \apj, 721, 193

\bibitem[{{P{\'e}rez-Gonz{\'a}lez} {et~al.}(2005){P{\'e}rez-Gonz{\'a}lez},
  {Rieke}, {Egami}, {Alonso-Herrero}, {Dole}, {Papovich}, {Blaylock}, {Jones},
  {Rieke}, {Rigby}, {Barmby}, {Fazio}, {Huang}, \&
  {Martin}}]{Perez-Gonzalez+2005}
{P{\'e}rez-Gonz{\'a}lez}, P.~G., {Rieke}, G.~H., {Egami}, E., {et~al.} 2005,
  \apj, 630, 82

\bibitem[{{Planck Collaboration} {et~al.}(2013){Planck Collaboration}, {Ade},
  {Aghanim}, {Armitage-Caplan}, {Arnaud}, {Ashdown}, {Atrio-Barandela},
  {Aumont}, {Baccigalupi}, {Banday}, \& et~al.}]{PC+13}
{Planck Collaboration}, {Ade}, P.~A.~R., {Aghanim}, N., {et~al.} 2013, ArXiv
  e-prints

\bibitem[{{Popesso} {et~al.}(2005){Popesso}, {Biviano}, {B{\"o}hringer},
  {Romaniello}, \& {Voges}}]{Popesso+05}
{Popesso}, P., {Biviano}, A., {B{\"o}hringer}, H., {Romaniello}, M., \&
  {Voges}, W. 2005, \aap, 433, 431

\bibitem[{{Popesso} {et~al.}(2014){Popesso}, {Biviano}, {Finoguenov}, {Wilman},
  {Salvato}, {Magnelli}, {Gruppioni}, {Pozzi}, {Rodighiero}, {Ziparo}, {Berta},
  {Elbaz}, {Dickinson}, {Lutz}, {Altieri}, {Aussel}, {Cimatti}, {Fadda},
  {Ilbert}, {Le Floch}, {Nordon}, {Poglitsch}, \& {Xu}}]{popesso+14}
{Popesso}, P., {Biviano}, A., {Finoguenov}, {et~al.} 2014, eprint
  arXiv:1407.8214

\bibitem[{{Popesso} {et~al.}(2012){Popesso}, {Biviano}, {Rodighiero},
  {Baronchelli}, {Salvato}, {Saintonge}, {Finoguenov}, {Magnelli}, {Gruppioni},
  {Pozzi}, {Lutz}, {Elbaz}, {Altieri}, {Andreani}, {Aussel}, {Berta}, {Capak},
  {Cava}, {Cimatti}, {Coia}, {Daddi}, {Dannerbauer}, {Dickinson}, {Dasyra},
  {Fadda}, {F{\"o}rster Schreiber}, {Genzel}, {Hwang}, {Kartaltepe}, {Ilbert},
  {Le Floch}, {Leiton}, {Magdis}, {Nordon}, {Patel}, {Poglitsch}, {Riguccini},
  {Sanchez Portal}, {Shao}, {Tacconi}, {Tomczak}, {Tran}, \&
  {Valtchanov}}]{popesso+12}
{Popesso}, P., {Biviano}, A., {Rodighiero}, G., {et~al.} 2012, \aap, 537, A58

\bibitem[{{Popesso} {et~al.}(2009){Popesso}, {Dickinson}, {Nonino}, {Vanzella},
  {Daddi}, {Fosbury}, {Kuntschner}, {Mainieri}, {Cristiani}, {Cesarsky},
  {Giavalisco}, {Renzini}, \& {GOODS Team}}]{Popesso+09}
{Popesso}, P., {Dickinson}, M., {Nonino}, M., {et~al.} 2009, \aap, 494, 443

\bibitem[{{Pratt} {et~al.}(2007){Pratt}, {B{\"o}hringer}, {Croston}, {Arnaud},
  {Borgani}, {Finoguenov}, \& {Temple}}]{pratt+07}
{Pratt}, G.~W., {B{\"o}hringer}, H., {Croston}, J.~H., {et~al.} 2007, \aap,
  461, 71

\bibitem[{{Prescott} {et~al.}(2006){Prescott}, {Impey}, {Cool}, \&
  {Scoville}}]{Prescott+06}
{Prescott}, M.~K.~M., {Impey}, C.~D., {Cool}, R.~J., \& {Scoville}, N.~Z. 2006,
  \apj, 644, 100

\bibitem[{{Reddy} {et~al.}(2008){Reddy}, {Steidel}, {Pettini}, {Adelberger},
  {Shapley}, {Erb}, \& {Dickinson}}]{reddy+08}
{Reddy}, N.~A., {Steidel}, C.~C., {Pettini}, M., {et~al.} 2008, \apjs, 175, 48

\bibitem[{{Rettura} {et~al.}(2010){Rettura}, {Rosati}, {Nonino}, {Fosbury},
  {Gobat}, {Menci}, {Strazzullo}, {Mei}, {Demarco}, \& {Ford}}]{Rettura+10}
{Rettura}, A., {Rosati}, P., {Nonino}, M., {et~al.} 2010, \apj, 709, 512

\bibitem[{{Robotham} {et~al.}(2011){Robotham}, {Norberg}, {Driver}, {Baldry},
  {Bamford}, {Hopkins}, {Liske}, {Loveday}, {Merson}, {Peacock}, {Brough},
  {Cameron}, {Conselice}, {Croom}, {Frenk}, {Gunawardhana}, {Hill}, {Jones},
  {Kelvin}, {Kuijken}, {Nichol}, {Parkinson}, {Pimbblet}, {Phillipps},
  {Popescu}, {Prescott}, {Sharp}, {Sutherland}, {Taylor}, {Thomas}, {Tuffs},
  {van Kampen}, \& {Wijesinghe}}]{Robotham+11}
{Robotham}, A.~S.~G., {Norberg}, P., {Driver}, S.~P., {et~al.} 2011, \mnras,
  416, 2640

\bibitem[{{Rosario} {et~al.}(2012){Rosario}, {Santini}, {Lutz}, {Shao},
  {Maiolino}, {Alexander}, {Altieri}, {Andreani}, {Aussel}, {Bauer}, {Berta},
  {Bongiovanni}, {Brandt}, {Brusa}, {Cepa}, {Cimatti}, {Cox}, {Daddi}, {Elbaz},
  {Fontana}, {F{\"o}rster Schreiber}, {Genzel}, {Grazian}, {Le Floch},
  {Magnelli}, {Mainieri}, {Netzer}, {Nordon}, {P{\'e}rez Garcia}, {Poglitsch},
  {Popesso}, {Pozzi}, {Riguccini}, {Rodighiero}, {Salvato}, {Sanchez-Portal},
  {Sturm}, {Tacconi}, {Valtchanov}, \& {Wuyts}}]{rosario+12}
{Rosario}, D.~J., {Santini}, P., {Lutz}, D., {et~al.} 2012, \aap, 545, A45

\bibitem[{{Rovilos} {et~al.}(2012){Rovilos}, {Comastri}, {Gilli},
  {Georgantopoulos}, {Ranalli}, {Vignali}, {Lusso}, {Cappelluti}, {Zamorani},
  {Elbaz}, {Dickinson}, {Hwang}, {Charmandaris}, {Ivison}, {Merloni}, {Daddi},
  {Carrera}, {Brandt}, {Mullaney}, {Scott}, {Alexander}, {Del Moro},
  {Morrison}, {Murphy}, {Altieri}, {Aussel}, {Dannerbauer}, {Kartaltepe},
  {Leiton}, {Magdis}, {Magnelli}, {Popesso}, \& {Valtchanov}}]{rovilos+12}
{Rovilos}, E., {Comastri}, A., {Gilli}, R., {et~al.} 2012, \aap, 546, A58

\bibitem[{{Rykoff} {et~al.}(2008){Rykoff}, {Evrard}, {McKay}, {Becker},
  {Johnston}, {Koester}, {Nord}, {Rozo}, {Sheldon}, {Stanek}, \&
  {Wechsler}}]{rykoff+08}
{Rykoff}, E.~S., {Evrard}, A.~E., {McKay}, T.~A., {et~al.} 2008, \mnras, 387,
  L28

\bibitem[{{Rykoff} {et~al.}(2012){Rykoff}, {Koester}, {Rozo}, {Annis},
  {Evrard}, {Hansen}, {Hao}, {Johnston}, {McKay}, \& {Wechsler}}]{Rykoff+12}
{Rykoff}, E.~S., {Koester}, B.~P., {Rozo}, E., {et~al.} 2012, \apj, 746, 178

\bibitem[{{Sanders} {et~al.}(2007){Sanders}, {Salvato}, {Aussel}, {Ilbert},
  {Scoville}, {Surace}, {Frayer}, {Sheth}, {Helou}, {Brooke}, {Bhattacharya},
  {Yan}, {Kartaltepe}, {Barnes}, {Blain}, {Calzetti}, {Capak}, {Carilli},
  {Carollo}, {Comastri}, {Daddi}, {Ellis}, {Elvis}, {Fall}, {Franceschini},
  {Giavalisco}, {Hasinger}, {Impey}, {Koekemoer}, {Le F{\`e}vre}, {Lilly},
  {Liu}, {McCracken}, {Mobasher}, {Renzini}, {Rich}, {Schinnerer}, {Shopbell},
  {Taniguchi}, {Thompson}, {Urry}, \& {Williams}}]{Sanders+07}
{Sanders}, D.~B., {Salvato}, M., {Aussel}, H., {et~al.} 2007, \apjs, 172, 86

\bibitem[{{Saunders} {et~al.}(1990){Saunders}, {Rowan-Robinson}, {Lawrence},
  {Efstathiou}, {Kaiser}, {Ellis}, \& {Frenk}}]{saunders+90}
{Saunders}, W., {Rowan-Robinson}, M., {Lawrence}, A., {et~al.} 1990, \mnras,
  242, 318

\bibitem[{{Schawinski} {et~al.}(2009){Schawinski}, {Virani}, {Simmons}, {Urry},
  {Treister}, {Kaviraj}, \& {Kushkuley}}]{schawinski+09}
{Schawinski}, K., {Virani}, S., {Simmons}, B., {et~al.} 2009, \apjl, 692, L19

\bibitem[{{Shao} {et~al.}(2010){Shao}, {Lutz}, {Nordon}, {Maiolino},
  {Alexander}, {Altieri}, {Andreani}, {Aussel}, {Bauer}, {Berta},
  {Bongiovanni}, {Brandt}, {Brusa}, {Cava}, {Cepa}, {Cimatti}, {Daddi},
  {Dominguez-Sanchez}, {Elbaz}, {F{\"o}rster Schreiber}, {Geis}, {Genzel},
  {Grazian}, {Gruppioni}, {Magdis}, {Magnelli}, {Mainieri}, {P{\'e}rez
  Garc{\'{\i}}a}, {Poglitsch}, {Popesso}, {Pozzi}, {Riguccini}, {Rodighiero},
  {Rovilos}, {Saintonge}, {Salvato}, {Sanchez Portal}, {Santini}, {Sturm},
  {Tacconi}, {Valtchanov}, {Wetzstein}, \& {Wieprecht}}]{Shao+10}
{Shao}, L., {Lutz}, D., {Nordon}, R., {et~al.} 2010, \aap, 518, L26

\bibitem[{{Silverman} {et~al.}(2010){Silverman}, {Mainieri}, {Salvato},
  {Hasinger}, {Bergeron}, {Capak}, {Szokoly}, {Finoguenov}, {Gilli}, {Rosati},
  {Tozzi}, {Vignali}, {Alexander}, {Brandt}, {Lehmer}, {Luo}, {Rafferty},
  {Xue}, {Balestra}, {Bauer}, {Brusa}, {Comastri}, {Kartaltepe}, {Koekemoer},
  {Miyaji}, {Schneider}, {Treister}, {Wisotski}, \& {Schramm}}]{silverman+10}
{Silverman}, J.~D., {Mainieri}, V., {Salvato}, M., {et~al.} 2010, \apjs, 191,
  124

\bibitem[{{Simha} {et~al.}(2009){Simha}, {Weinberg}, {Dav{\'e}}, {Gnedin},
  {Katz}, \& {Kere{\v s}}}]{simha+09}
{Simha}, V., {Weinberg}, D.~H., {Dav{\'e}}, R., {et~al.} 2009, \mnras, 399, 650

\bibitem[{{Smail} {et~al.}(2014){Smail}, {Geach}, {Swinbank}, {Tadaki},
  {Arumugam}, {Hartley}, {Almaini}, {Bremer}, {Chapin}, {Chapman}, {Danielson},
  {Edge}, {Scott}, {Simpson}, {Simpson}, {Conselice}, {Dunlop}, {Ivison},
  {Karim}, {Kodama}, {Mortlock}, {Robson}, {Roseboom}, {Thomson}, {van der
  Werf}, \& {Webb}}]{Smail+14}
{Smail}, I., {Geach}, J.~E., {Swinbank}, A.~M., {et~al.} 2014, \apj, 782, 19

\bibitem[{{Soifer} {et~al.}(2008){Soifer}, {Helou}, \& {Werner}}]{soifer+08}
{Soifer}, B.~T., {Helou}, G., \& {Werner}, M. 2008, \araa, 46, 201

\bibitem[{{Springel} {et~al.}(2005){Springel}, {White}, {Jenkins}, {Frenk},
  {Yoshida}, {Gao}, {Navarro}, {Thacker}, {Croton}, {Helly}, {Peacock}, {Cole},
  {Thomas}, {Couchman}, {Evrard}, {Colberg}, \& {Pearce}}]{springel+05}
{Springel}, V., {White}, S.~D.~M., {Jenkins}, A., {et~al.} 2005, \nat, 435, 629

\bibitem[{{Sun}(2012)}]{sun12}
{Sun}, M. 2012, New Journal of Physics, 14, 045004

\bibitem[{{Tacconi} {et~al.}(2010){Tacconi}, {Genzel}, {Neri}, {Cox}, {Cooper},
  {Shapiro}, {Bolatto}, {Bouch{\'e}}, {Bournaud}, {Burkert}, {Combes},
  {Comerford}, {Davis}, {Schreiber}, {Garcia-Burillo}, {Gracia-Carpio}, {Lutz},
  {Naab}, {Omont}, {Shapley}, {Sternberg}, \& {Weiner}}]{tacconi+10}
{Tacconi}, L.~J., {Genzel}, R., {Neri}, R., {et~al.} 2010, \nat, 463, 781

\bibitem[{{Tanaka} {et~al.}(2013){Tanaka}, {Finoguenov}, {Mirkazemi}, {Wilman},
  {Mulchaey}, {Ueda}, {Xue}, {Brandt}, \& {Cappelluti}}]{tanaka+13}
{Tanaka}, M., {Finoguenov}, A., {Mirkazemi}, M., {et~al.} 2013, \pasj, 65, 17

\bibitem[{{Tinker} {et~al.}(2008){Tinker}, {Kravtsov}, {Klypin}, {Abazajian},
  {Warren}, {Yepes}, {Gottl{\"o}ber}, \& {Holz}}]{Tinker+08}
{Tinker}, J., {Kravtsov}, A.~V., {Klypin}, A., {et~al.} 2008, \apj, 688, 709

\bibitem[{{Trump} {et~al.}(2007){Trump}, {Impey}, {McCarthy}, {Elvis},
  {Huchra}, {Brusa}, {Hasinger}, {Schinnerer}, {Capak}, {Lilly}, \&
  {Scoville}}]{Trump+07}
{Trump}, J.~R., {Impey}, C.~D., {McCarthy}, P.~J., {et~al.} 2007, \apjs, 172,
  383

\bibitem[{{Vale} \& {Ostriker}(2004)}]{Vale+04}
{Vale}, A. \& {Ostriker}, J.~P. 2004, \mnras, 353, 189

\bibitem[{{van de Voort} {et~al.}(2011){van de Voort}, {Schaye}, {Booth}, \&
  {Dalla Vecchia}}]{vandevoort+11}
{van de Voort}, F., {Schaye}, J., {Booth}, C.~M., \& {Dalla Vecchia}, C. 2011,
  \mnras, 415, 2782

\bibitem[{{Vanzella} {et~al.}(2006){Vanzella}, {Cristiani}, {Dickinson},
  {Kuntschner}, {Nonino}, {Rettura}, {Rosati}, {Vernet}, {Cesarsky},
  {Ferguson}, {Fosbury}, {Giavalisco}, {Grazian}, {Haase}, {Moustakas},
  {Popesso}, {Renzini}, {Stern}, \& {GOODS Team}}]{vanzella+06}
{Vanzella}, E., {Cristiani}, S., {Dickinson}, M., {et~al.} 2006, \aap, 454, 423

\bibitem[{{Verdes-Montenegro} {et~al.}(2001){Verdes-Montenegro}, {Yun},
  {Williams}, {Huchtmeier}, {Del Olmo}, \& {Perea}}]{verdes+01}
{Verdes-Montenegro}, L., {Yun}, M.~S., {Williams}, B.~A., {et~al.} 2001, \aap,
  377, 812

\bibitem[{{Wang} {et~al.}(2007){Wang}, {Yang}, {Mo}, \& {van den
  Bosch}}]{wang+07}
{Wang}, Y., {Yang}, X., {Mo}, H.~J., \& {van den Bosch}, F.~C. 2007, \apj, 664,
  608

\bibitem[{{Wetzel} {et~al.}(2013){Wetzel}, {Tinker}, {Conroy}, \& {van den
  Bosch}}]{wetzel+13}
{Wetzel}, A.~R., {Tinker}, J.~L., {Conroy}, C., \& {van den Bosch}, F.~C. 2013,
  \mnras, 432, 336

\bibitem[{{Wild} {et~al.}(2010){Wild}, {Heckman}, \& {Charlot}}]{wild+10}
{Wild}, V., {Heckman}, T., \& {Charlot}, S. 2010, \mnras, 405, 933

\bibitem[{{Williams} {et~al.}(2012){Williams}, {Kelson}, {Mulchaey},
  {Dressler}, {McCarthy}, \& {Shectman}}]{williams+12}
{Williams}, R.~J., {Kelson}, D.~D., {Mulchaey}, J.~S., {et~al.} 2012, \apjl,
  749, L12

\bibitem[{{Yang} {et~al.}(2007){Yang}, {Mo}, {van den Bosch}, {Pasquali}, {Li},
  \& {Barden}}]{Yang+07}
{Yang}, X., {Mo}, H.~J., {van den Bosch}, F.~C., {et~al.} 2007, \apj, 671, 153

\bibitem[{{Yang} {et~al.}(2012){Yang}, {Mo}, {van den Bosch}, {Zhang}, \&
  {Han}}]{Yang+12}
{Yang}, X., {Mo}, H.~J., {van den Bosch}, F.~C., {Zhang}, Y., \& {Han}, J.
  2012, \apj, 752, 41

\bibitem[{{Yesuf} {et~al.}(2014){Yesuf}, {Faber}, {Trump}, {Koo}, {Fang},
  {Liu}, {Wild}, \& {Hayward}}]{yesuf+14}
{Yesuf}, H.~M., {Faber}, S.~M., {Trump}, J.~R., {et~al.} 2014, \apj, 792, 84

\bibitem[{{Zabludoff} \& {Mulchaey}(1998)}]{ZM98let}
{Zabludoff}, A.~I. \& {Mulchaey}, J.~S. 1998, \apjl, 498, L5

\bibitem[{{Ziparo} {et~al.}(2013){Ziparo}, {Popesso}, {Biviano}, {Finoguenov},
  {Wuyts}, {Wilman}, {Salvato}, {Tanaka}, {Ilbert}, {Nandra}, {Lutz}, {Elbaz},
  {Dickinson}, {Altieri}, {Aussel}, {Berta}, {Cimatti}, {Fadda}, {Genzel}, {Le
  Flo'ch}, {Magnelli}, {Nordon}, {Poglitsch}, {Pozzi}, {Portal}, {Tacconi},
  {Bauer}, {Brandt}, {Cappelluti}, {Cooper}, \& {Mulchaey}}]{ziparo+13}
{Ziparo}, F., {Popesso}, P., {Biviano}, A., {et~al.} 2013, \mnras, 434, 3089

\bibitem[{{Ziparo} {et~al.}(2014){Ziparo}, {Popesso}, {Finoguenov}, {Biviano},
  {Wuyts}, {Wilman}, {Salvato}, {Tanaka}, {Nandra}, {Lutz}, {Elbaz},
  {Dickinson}, {Altieri}, {Aussel}, {Berta}, {Cimatti}, {Fadda}, {Genzel}, {Le
  Floc'h}, {Magnelli}, {Nordon}, {Poglitsch}, {Pozzi}, {Portal}, {Tacconi},
  {Bauer}, {Brandt}, {Cappelluti}, {Cooper}, \& {Mulchaey}}]{Ziparo+14}
{Ziparo}, F., {Popesso}, P., {Finoguenov}, A., {et~al.} 2014, \mnras, 437, 458

\end{thebibliography}

\end{document}